\documentclass[           
               showpacs,            
               preprintnumbers,     
               aps,                 
               prd,                 
               superscriptaddress,      
               nofootinbib,         
               tightenlines,        
               floats,floatfix      
               ]{revtex4}

\usepackage{amsmath}
\usepackage{amssymb}
\usepackage{graphicx}

\begin{document}


\title{On Wormholes supported by phantom energy}


\author{J. A. Gonz\'alez}
\affiliation{Instituto de F\'{\i}sica y Matem\'{a}ticas, Universidad
              Michoacana de San Nicol\'as de Hidalgo. Edificio C-3, Cd.
              Universitaria, A. P. 2-82, 58040 Morelia, Michoac\'{a}n,
              M\'{e}xico.}

\author{F. S. Guzm\'an}
\affiliation{Instituto de F\'{\i}sica y Matem\'{a}ticas, Universidad
              Michoacana de San Nicol\'as de Hidalgo. Edificio C-3, Cd.
              Universitaria, A. P. 2-82, 58040 Morelia, Michoac\'{a}n,
              M\'{e}xico.}

\author{N. Montelongo-Garc\'ia}
\affiliation{Instituto de F\'{\i}sica y Matem\'{a}ticas, Universidad
              Michoacana de San Nicol\'as de Hidalgo. Edificio C-3, Cd.
              Universitaria, A. P. 2-82, 58040 Morelia, Michoac\'{a}n,
              M\'{e}xico.}

\author{T. Zannias}
\affiliation{Instituto de F\'{\i}sica y Matem\'{a}ticas, Universidad
              Michoacana de San Nicol\'as de Hidalgo. Edificio C-3, Cd.
              Universitaria, A. P. 2-82, 58040 Morelia, Michoac\'{a}n,
              M\'{e}xico.}


\date{\today}


\begin{abstract}

\noindent By a combination of analytical and numerical techniques,
we demonstrate the existence of spherical, asymptotically flat
traversable wormholes supported by exotic matter whose stress tensor
relative to the orthonormal frame of Killing observers takes the
form of a perfect fluid possessing anisotropic pressures and subject
to linear equations of state: $\tau=\lambda\rho c^{2}$, $P=\mu\rho
c^{2}$. We show that there exists a four
parameter family of asymptotically flat spherical wormholes
parametrized by the  area of the
throat $A(0)$, the gradient $\Lambda(0)$ of the red shift factor 
evaluated at the throat as well as the  values of $(\lambda, \mu)$.
The latter are subject to restrictions: $\lambda>1$ and  $2\mu>\lambda$ or
$\lambda<0$ and  $2\mu<-|\lambda|$. For particular values of $(\lambda, \mu)$,
the stress tensor may be interpreted as representing a phantom
configuration, while for
other values represents exotic matter. All
solutions  have the property that the two asymptotically flat ends
posses finite ADM mass.
\end{abstract}




\maketitle



\section{Introduction\label{I}}

\noindent The observed late time accelerated expansion of the
universe \cite{Riess,Netterfield} appears to imply that the universe
is dominated by dark energy \cite{C-L-Bennett}, a substance having
the property that if  $(\rho c^{2}, P)$ stand for the energy density
and isotropic pressure as measured by observes co-moving with the
expansion, then $P$ is negative or more precisely $\rho c^{2}
+3P<0$. This property is often stated that the equation of state
(EOS) for dark energy is $w=\frac{P}{\rho c^{2}}<-\frac{1}{3}.$
Additional observations
 of CMB, gravitational lensing and supernovae \cite{A-G-Riess},
mildly favor a more exotic EOS: $w=\frac{P}{\rho c^2}<-1$ and such
substance is dubbed as phantom energy. The possibility that the
cosmic fluid is in state of phantom energy
 has another consequence. The
stress tensor describing phantom energy violates the null energy
condition and thus phantom energy is the type 
matter required to support wormholes \cite{Morris}.

The issue whether a static spherical distribution of phantom energy
can support spherical wormholes has been addressed in \cite{Sushkov,Lobo}.
 In these works phantom 
energy is modeled by a
perfect fluid stress tensor possessing
 anisotropic pressures with the
radial pressure $P_{r}=-\tau$  and energy density $\rho c^{2}$
obeying: $w=\frac{P_{r}}{\rho c^2}=-\frac{\tau}{\rho c^{2}}<-1$
while the tangential pressure $P$  is defined by the field equations. In
\cite{Sushkov} an a-priori choice for the energy density has been
made  whereas in \cite{Lobo} the wormhole metric has been a priori
specified.

In the present work, we model exotic matter by a stress tensor so
that relative to orthonormal frame  of Killing observers takes the
form of a perfect fluid  possessing
anisotropic pressures but we assume that $\rho
c^{2}$, $ \tau$ and $P$  are subject
to constraints referred as the EOS. Guided by the cosmological
phantom EOS and generality purposes we adopt \cite{N2}:

\begin{eqnarray}
\tau=\lambda \rho c^{2},~~~
P=\mu \rho c^{2},~~~
\lambda,\mu ~ \in ~ I\!\!R \label{J},
\end{eqnarray}

\noindent
where $(\lambda,\mu)$ are treated as free parameters subject only to the
restriction that the stress tensor should violate the null energy
condition in an open vicinity of the throat and
restrictions arising by demanding asymptotic flatness. The problem of
the existence of
spherical wormholes supported by this type of exotic matter, is
formulated as an initial value problem (IVP here after) with the
throat serving as initial value surface (for more details regarding
this approach see \cite{MZ1}, \cite{MZ2}). By a combination of
analytical and numerical techniques we show
that there exist initial conditions and values
of $(\lambda,\mu)$ where solutions of this IVP describe
asymptotically flat non singular wormholes. Asymptotic flatness
requires $(\lambda,\mu)$ to obey either
$\lambda>1$ and $2\mu>\lambda$ or $\lambda<0$ and $2\mu<-|\lambda|$.
Thus  even though
$(\lambda,\mu)$ play a secondary role in the behavior of the
solutions near the throat, they determine the asymptotic behavior of
the solution.

The structure of the present paper is as follows: In the following
section we formulate the relevant IVP and discuss some properties of this IVP
while in sections \ref{III} we
discuss numerical solutions. In 
section \ref{IV} we present an analysis that explains the
numerical outputs and advance arguments that set restrictions upon
$(\lambda,\mu)$ so that the solutions describe  asymptotically flat
wormholes. We finish the paper with a discussion concerning the
results and future work.


\section{Local wormholes supported by a ``perfect fluid''\label{II}}

\noindent We consider a static, spherically symmetric wormhole:

\begin{eqnarray}
\textbf{g}=-e^{2\Phi(l)}dt^{2}+dl^2+r^{2}(l)d\Omega^2,
~~ l ~ \in ~(-\alpha,\alpha),~~\alpha>0,\label{p}
\end{eqnarray}

\noindent where $r(l=0)\neq 0$, $\frac{dr(l)}{dl}|_{0}=0,$
$\frac{d^2r(l)}{dl^2}|_{0}>0$ and  thus $l=0$ marks the location of
the throat. We support this wormhole with a stress tensor
$T_{\alpha\beta}$ so that relative to the orthonormal frame of
Killing observers decomposes according to:

\begin{eqnarray}
T_{\alpha\beta}&=&\rho(l)
    c^2u_{\alpha}u_{\beta}-\tau(l)X_{\alpha}X_{\beta}+
    P(l)(Y_{\alpha}Y_{\beta}+Z_{\alpha}Z_{\beta}),\label{y1}\\
\nonumber \textbf{u}&=&e^{-\Phi(l)}\frac{\partial}{\partial t},
X=\frac{\partial}{\partial l},\,\, Y=\frac{1}{r(l)}\frac{\partial}
{\partial\theta},\,\,\,Z=\frac{1}{r(l)\sin\theta}
\frac{\partial}{\partial\phi}.
\end{eqnarray}

In the gauge of (\ref{p}), if we define $\hat{r}(l)={r}^{-1}(l)$,
introduce
 $K(l)=\frac{2}{r(l)}\frac{dr(l)}{dl}$ the trace
of the extrinsic curvature of the $SO(3)$ orbits as embedded in
$t=$cons hyper-surfaces and $\Lambda(l)=\frac{d\Phi(l)}{dl}$, then
$G_{\alpha\beta}=\hat{k}T_{\alpha\beta}$ taking into account
(\ref{J}) can be cast in the form \cite{MZ1,MZ2}:

\begin{eqnarray}
&&\frac{d\hat{r}(l)}{dl}=-\frac{1}{2}K(l)\hat{r}(l),\label{a3}\\
&&\frac{d K(l)}{dl}=-\frac{3}{4}K^2(l)+\hat{r}^2(l)
                     -\hat{k}\rho(l)c^2,\label{a4}\\
&&\frac{d\Lambda(l)}{dl}=-K(l)\Lambda(l)
                          -\Lambda^2(l)
                        +(1-\lambda+2\mu)\frac{\hat{k}\rho(l)c^{2}}{2},\label{a5}\\
&&\frac{d\rho(l)}{dl}=\rho(l)\left[(\frac{1}{\lambda}-1)\Lambda(l)
                     -(1+\frac{\mu}{\lambda})K(l)\right],\label{a6}\\
&&-\hat{r}^{2}(l)+\frac{K^{2}(l)}{4}
+\hat{k}\rho(l)c^{2}\lambda+\Lambda(l)K(l)=0.\label{a7}
\end{eqnarray}

A throat of area  $A(0)=4\pi r^{2}(0)$ requires tension $\tau(0)$ so
that $\hat{k}\tau(0)=r^{-2}(0)$ and moreover:
$(\tau(0)-\rho(0)c^2)>0$ \cite{note1}. These conditions in  view of (\ref{J})
take the form:

\begin{eqnarray}
\hat{k}\lambda c^{2}\rho
(0)=r^{-2}(0),~~~[\tau(0)-\rho(0)c^{2}]=(\lambda-1)\rho(0)c^{2}>0,
\label{F}
\end{eqnarray}

\noindent and thus they hold, provided either  $(\rho(0)>0, \lambda>1) $ or
$(\rho(0)<0, \lambda<0)$. If we adopt as a definition that an
inhomogeneous phantom configuration satisfies $w=\frac{P_{r}}{\rho
c^{2}}=-\frac{\tau}{\rho c^{2}}<-1$, then  the first choice yields:
$w=-\lambda<-1$ and thus $T_{\alpha\beta}$ describes a phantom like
configuration at least in an open vicinity of the throat. For the
second choice, even though $\rho(0)<0$ nevertheless $\tau(0)$ is
positive and for this case we interpret $T_{\alpha\beta}$ as
describing exotic matter \cite{N3}.

Any solution of (\ref{a3}-\ref{a7}) describes a local wormhole
having a throat of area $A(0)=4\pi r^{2}(0)=4\pi\hat{r}^{-2}(0)$ at
a prescribed $\Lambda(0)$, provided it satisfies the initial
conditions \cite{note1}:

\begin{eqnarray}
\hat{r}(0)=\left(\frac{4\pi}{A(0)}\right)^{1/2},
~~K(0)=0,~~\Lambda(0),~~\rho(0)=
\frac{4\pi}{\lambda\hat{k}c^{2}A(0)}.\label{X}
\end{eqnarray}

\noindent The system (\ref{a3}-\ref{a6}) combined with these initial
conditions constitutes  a well defined IVP. The theorem of
``Picard-Lindelof''  \cite{Picard} assures the local existence of a
unique $C^{1}$ solution defined on $[-b,b]\subset(-\alpha,\alpha)$.
Ideally, we would like to find necessary or (and) sufficient
conditions so that these local solutions are extendible for all
$l~\in~(-\infty,\infty)$ and moreover represent
asymptotically flat wormholes. Due to the non linearities in
(\ref{a3}-\ref{a6}) this is a formidable task. In this work we shall
gain insights into the properties of the maximal solutions  by first
resorting to numerical techniques. It is our hope that these
insights  will eventually contribute towards an analytical treatment of the 
problem.

For the purpose of numerical integrations, we compactify the
coordinate $l$ via:

\begin{eqnarray}
l(x)=\frac{r_{0}x}{1-x^{2}},~~~ x \in (-1,1), ~~r_{0}=r(0). \label{Tra}
\end{eqnarray}

\noindent This  transformation maps  $l\to\pm \infty$  into $x\to\pm\ 1$, $l=0$ into $x=0$ and transforms
(\ref{a3}-\ref{a6}) into:

\begin{eqnarray}
(1-x^{2})^{2}\frac{d\hat{r}(x)}{dx}
    &=&-\frac{r_{0}}{2}(1+x^{2})K(x)\hat{r}(x),\label{a}\\
(1-x^{2})^{2}\frac{dK(x)}{dx}
&=&r_{0}(1+x^{2})\left[-\frac{3}{4}K^{2}(x)
    +\hat{r}^{2}(x)-\hat{k}\rho(x)c^{2}\right],\label{b}\\
(1-x^{2})^{2}\frac{d\Lambda(x)}{dx}
    &=&r_{0}(1+x^{2})\left[-K(x)\Lambda(x)-
    \Lambda^{2}(x) + \frac{\hat{k}\rho(x)c^{2}}{2}
    (1-\lambda+2\mu)\right],\label{c}\\
(1-x^{2})^{2}\frac{d\rho(x)c^{2}}{dx}
    &=&r_{0}(1+x^{2})\left[\left(\frac{1}{\lambda}-1\right)\Lambda(x)
    -\left(1+\frac{\mu}{\lambda}\right)K(x)\right]\rho(x)c^{2},
\label{eq:final}
\end{eqnarray}

\noindent and leaves the initial conditions (\ref{X}) form invariant. 
We integrate the IVP defined by these equations 
combined with (\ref{X}) for
a variety of initial conditions
and values of $(\lambda,\mu)$. Before we enter into numerics,
we mention three properties of smooth solutions of this IVP.

\begin{itemize}

 \item Any $C^{1}$ solution $(\hat{r}(x),K(x),\Lambda(x),\rho(x))$
generated by initial conditions so that $\Lambda(0)=0$,
is ``reflectionally symmetric'' relative to the throat i.e. under ``
parity'' transformation $x\to -x$ behaves  according to \cite{XXX}:

\begin{eqnarray}
r(x)=r(-x),~ K(x)=-K(-x),~\Lambda(x)=-\Lambda(-x),~\rho(x)=\rho(-x). \label{B32}
\end{eqnarray}

\item  If  $(\hat{r}(x),K(x),\Lambda(x),\rho(x))$ is any $C^{1}$
solution generated by an arbitrary set of initial conditions, then the
functions:

\begin{eqnarray}
R(x)=\hat{r}(-x),~K_{1}(x)=-K(-x),L(x)=-\Lambda(-x),~\hat{\rho}(x)=\rho(-x),\label{B33}
\end{eqnarray}

define a new solution that satisfies the same initial
conditions as $(\hat{r}(x),K(x),\Lambda(x),\rho(x))$ does, except that
$L(0)=-\Lambda(0)$.

\item If $(\hat{r}(x),K(x),\Lambda(x),\rho(x))$ is any solution,
then under rescaling of the throat

\begin{eqnarray}
\hat{r}(0) \to\hat{r}^{'} (0)=A\hat{r}(0),~~A>0,
\end{eqnarray}

it follows that

\begin{eqnarray}
 A\hat{r}(x), ~~ AK(x), ~~ A\Lambda(x), ~~ A^{2}\rho(x), \label{B34}
\end{eqnarray}

satisfy (\ref{a}-\ref{eq:final}) and the rescaled initial
conditions (\ref{X}).

\end{itemize}

\noindent
Property (\ref{B32}) implies that for
reflectionally symmetric solutions, integrating
(\ref{a}-\ref{eq:final}) on the domain  $(0,1)$ would be sufficient.
Property (\ref{B33}) allows us to concentrate on solutions obeying
$\Lambda(0)\geq 0$ while (\ref{B34}) permits  us  to set the throat
radius at some convenient value.


\section{Numerical results\label{III}}

\noindent We integrate (\ref{a}-\ref{eq:final}) subject to (\ref{X}),
using a Runge-Kutta integrator of different accuracy and in order
to avoid the singularity
at $x= \pm 1$, we stagger the exact points $x=\pm 1$.
In all runs we employ units so that  $c=G=1$ and thus
$\hat{k}=8\pi$. We set the throat radius $r(0)=1$
and employ the following two sets of initial conditions \cite{note1}:

\begin{eqnarray}
&&\hat{r}(0)=1,\,\,\,K(0)=0,~ ~\Lambda(0),~
~\rho(0)=\frac{1}{8\pi\lambda},~~\lambda>1,\label{C1}\\
&&\hat{r}(0)=1,\,\,\,K(0)=0,~ ~\Lambda(0),~
~\rho(0)=\frac{1}{8\pi\lambda},~~\lambda<0 \label{C2}.
\end{eqnarray}

\begin{figure}[ht]
\includegraphics[width=8cm]{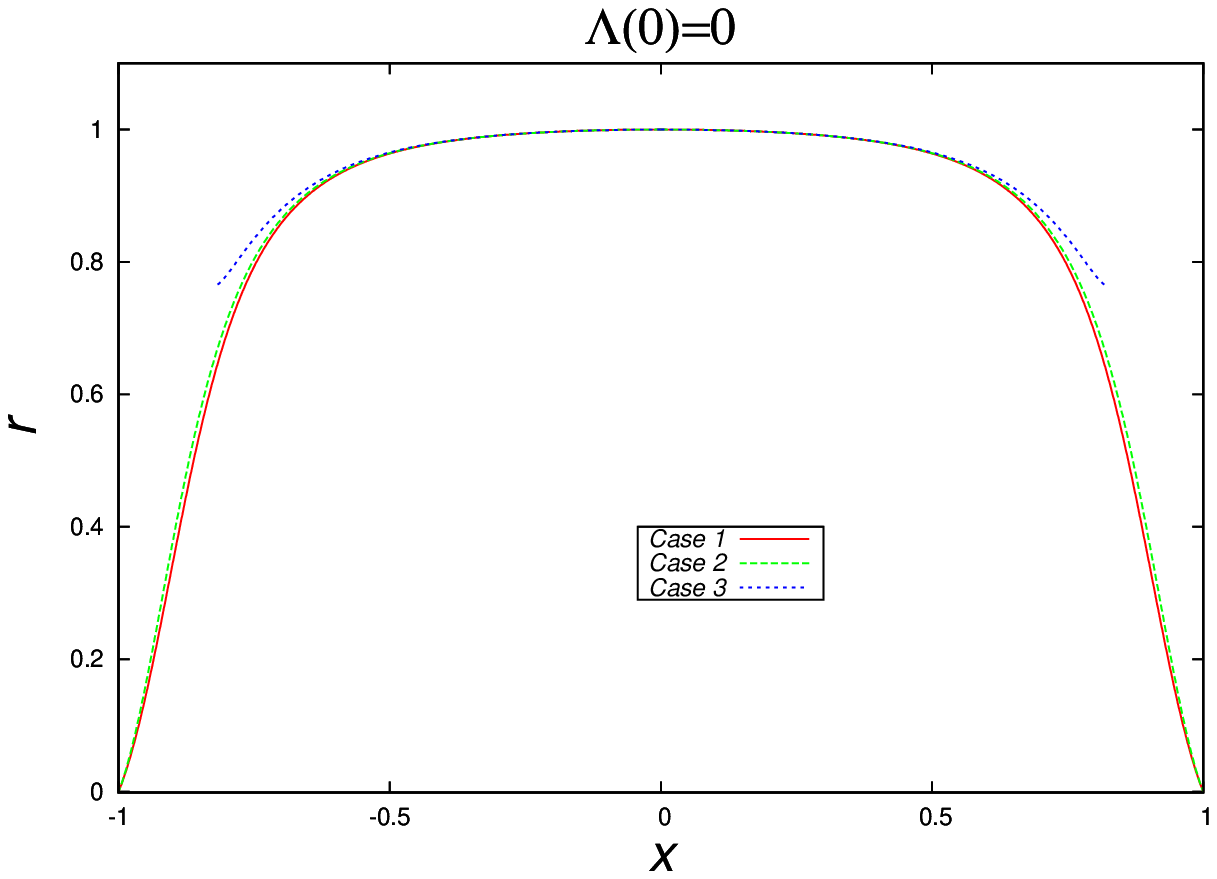}
\includegraphics[width=8cm]{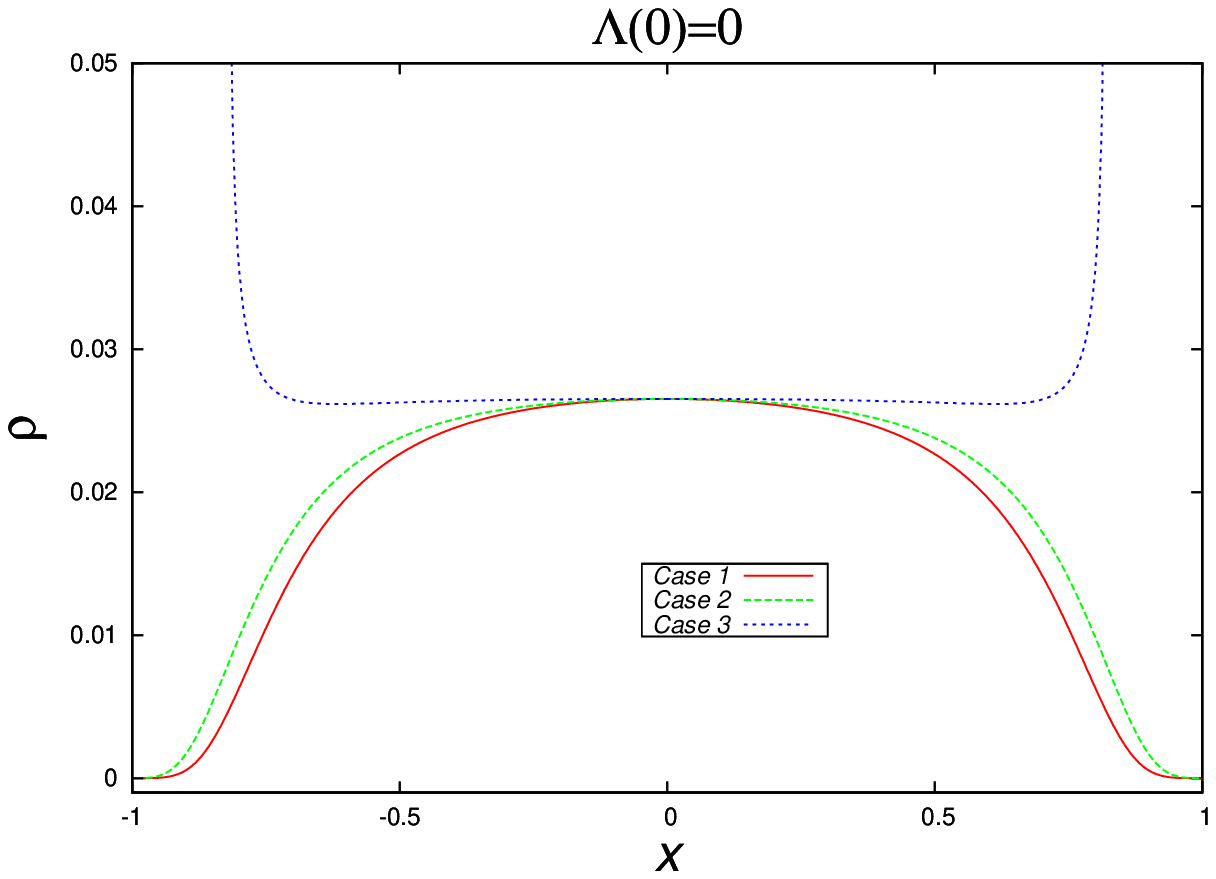}
\includegraphics[width=8cm]{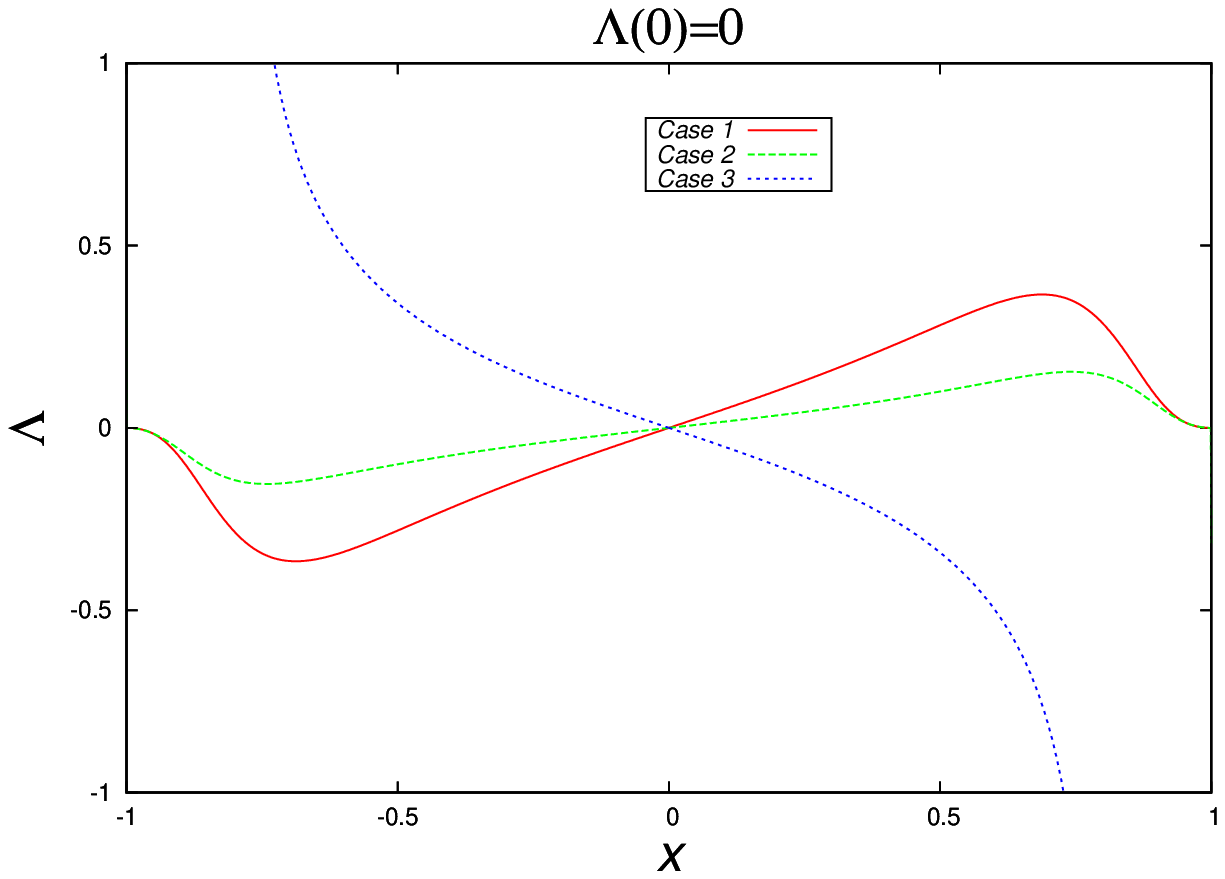}
\includegraphics[width=8cm]{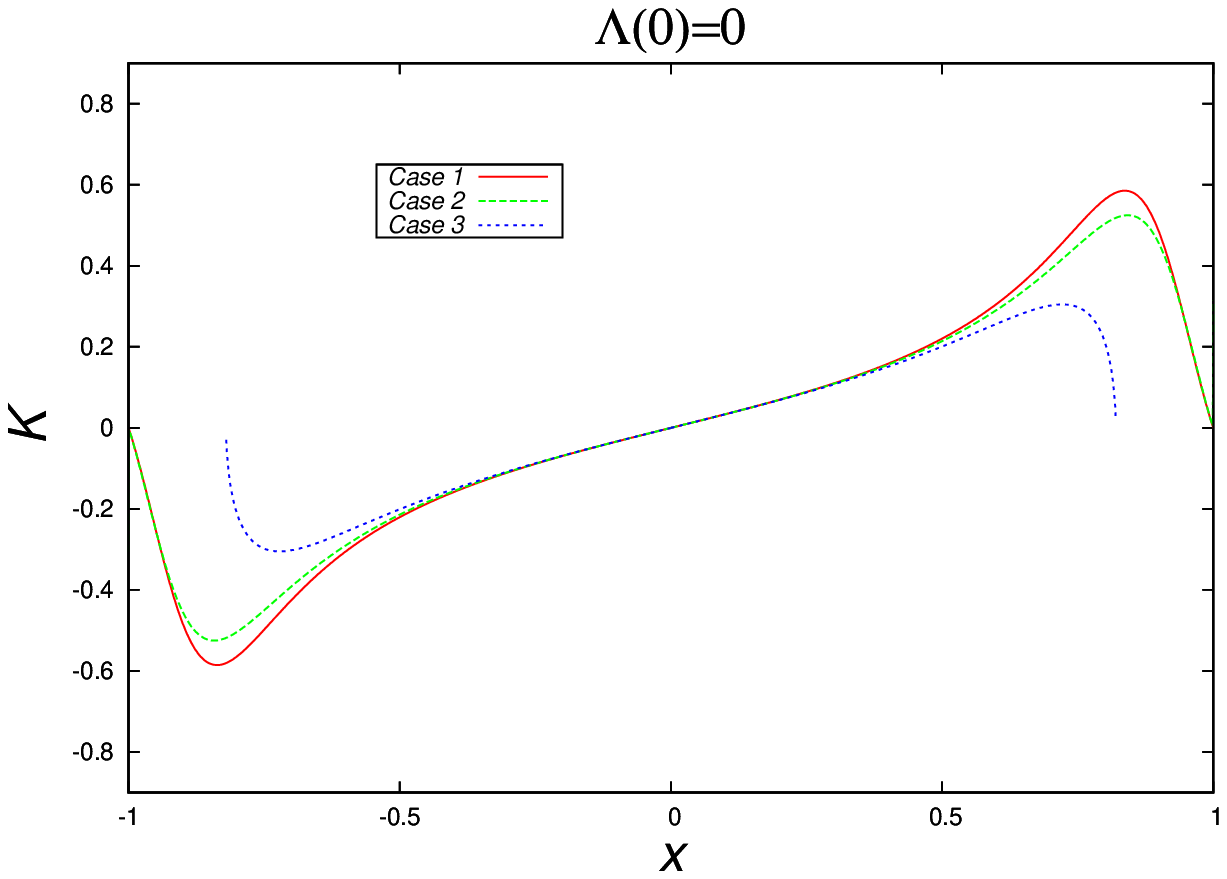}
\caption{Reflectionally symmetric solutions generated choosing:
$\Lambda(0)=0$ and  $(\lambda, \mu) $ labeled as case
(1,2,3) in Table I.)
 \label{figure1}}
\end{figure}

\begin{figure}[ht]
\includegraphics[width=8cm]{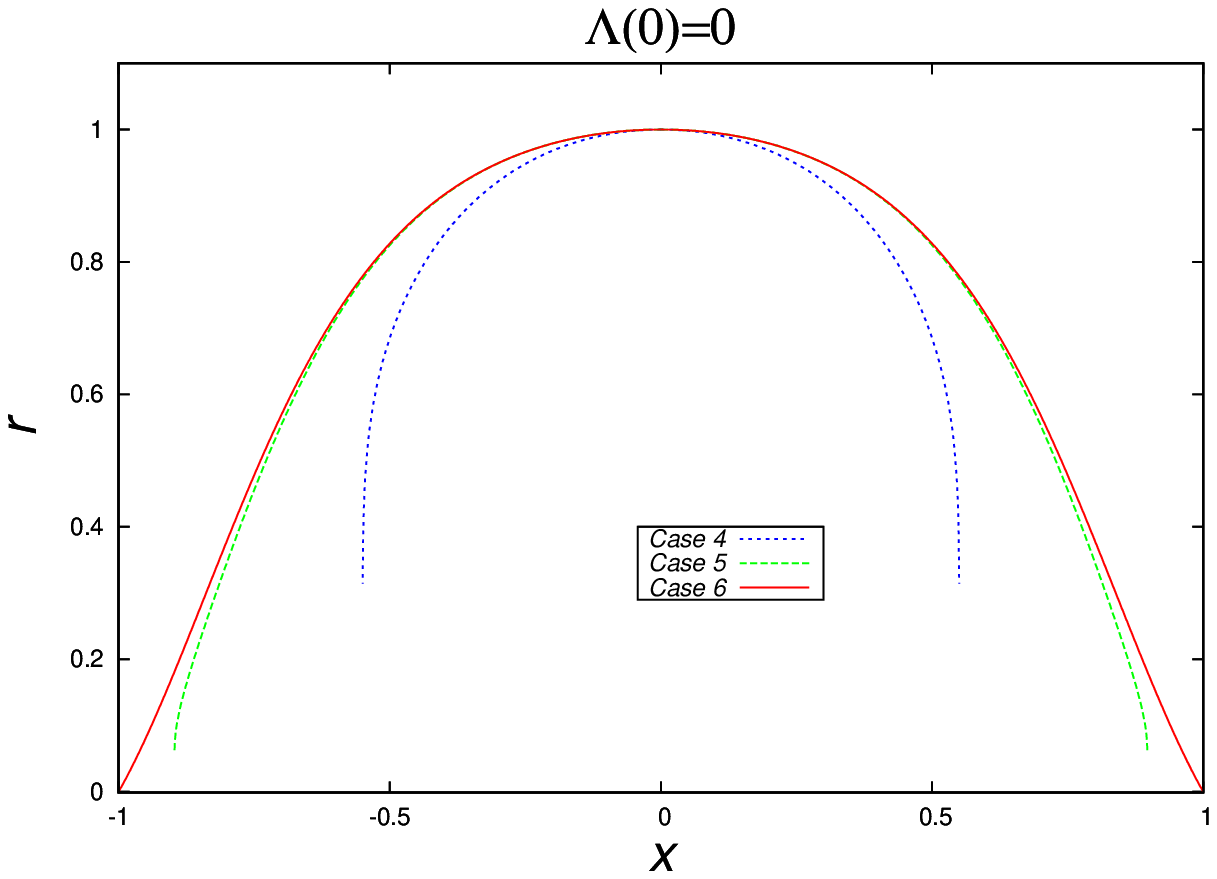}
\includegraphics[width=8cm]{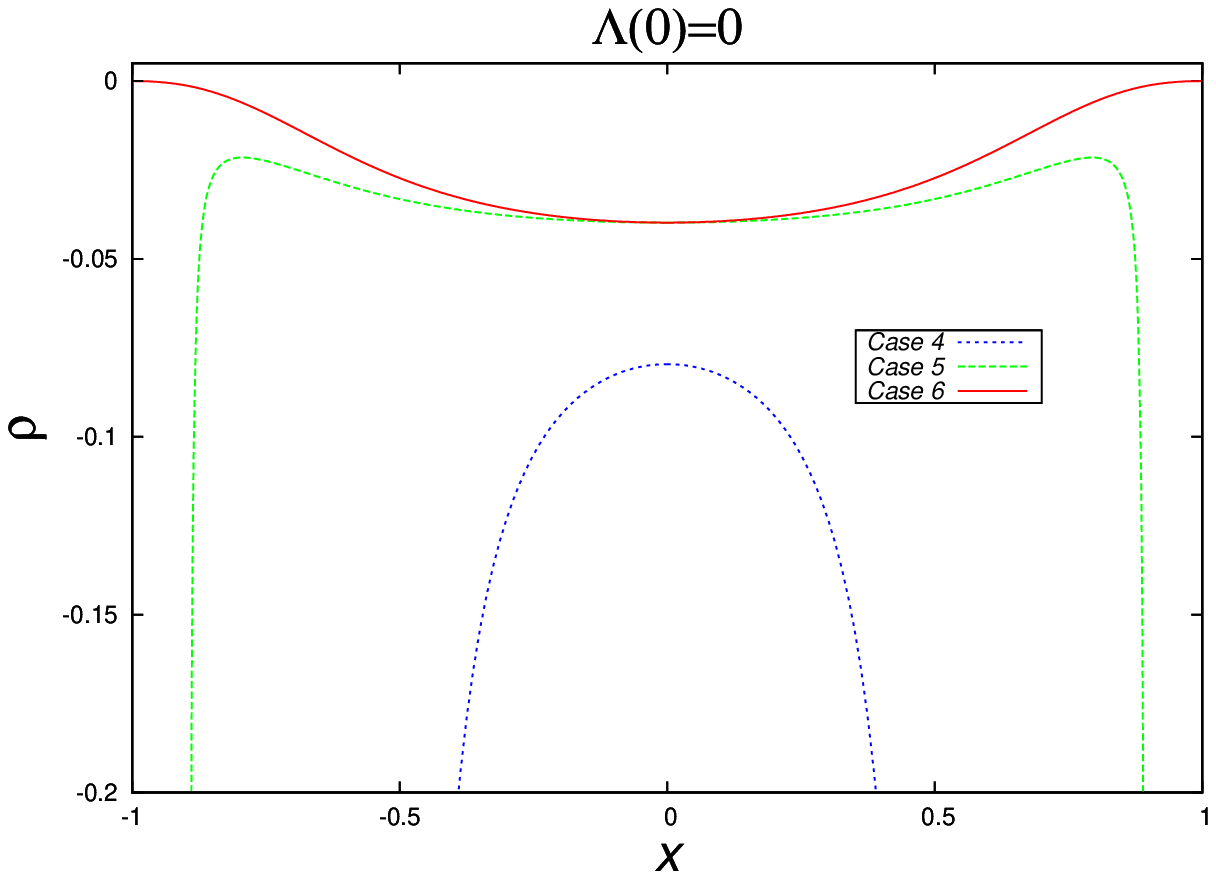}
\includegraphics[width=8cm]{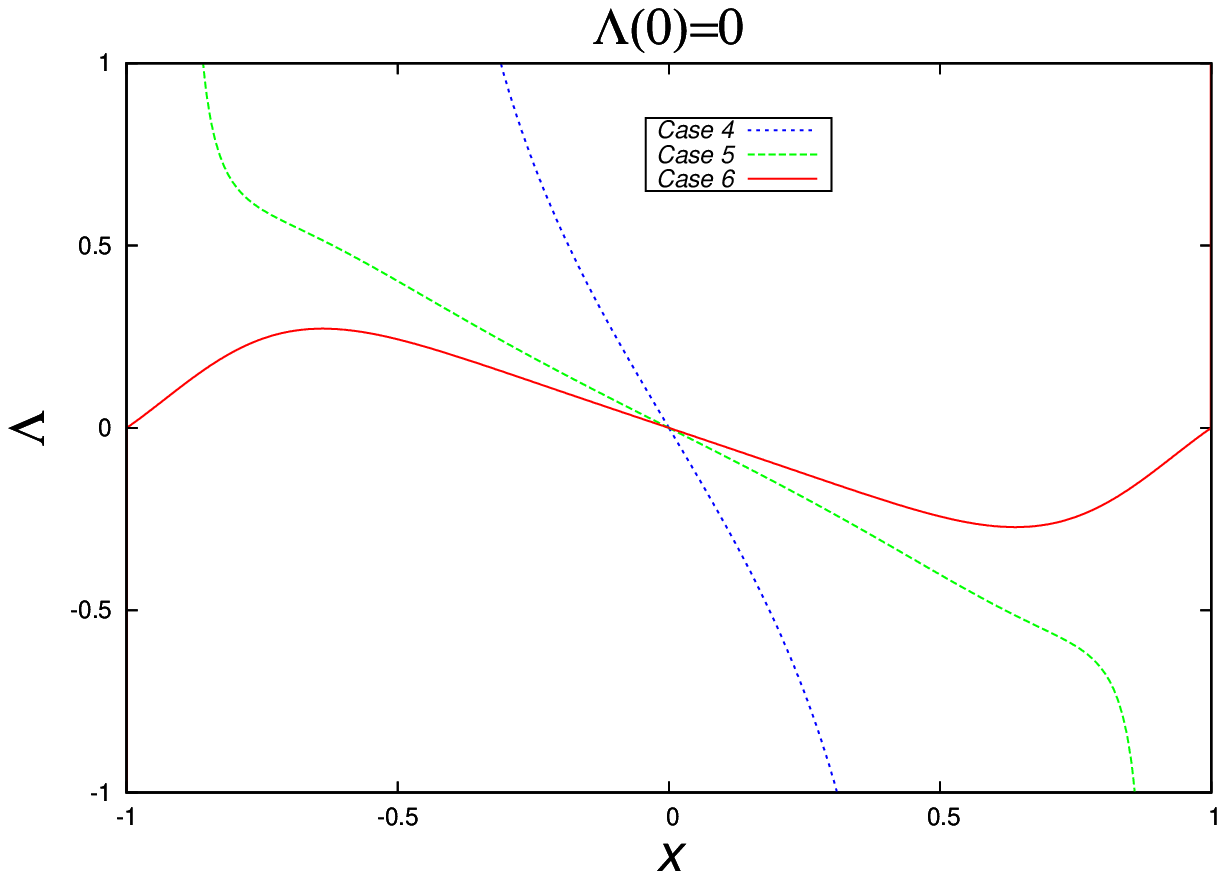}
\includegraphics[width=8cm]{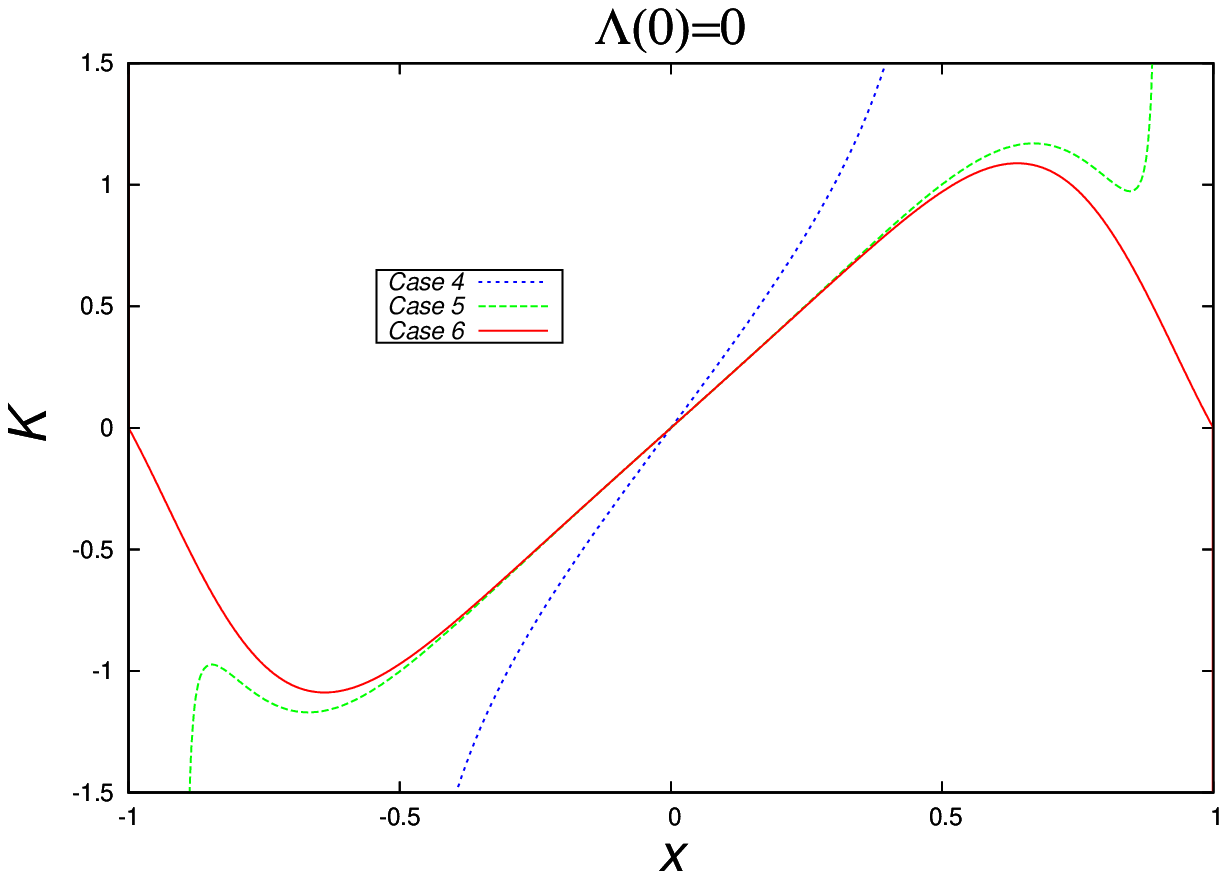}
\caption{Reflectionally symmetric solutions generated by choosing:
$\Lambda(0)=0,$ and  $(\lambda, \mu) $ labeled as case
(4,5,6) in Table.) \label{figure2}}
\end{figure}

\noindent At first we consider reflectionally symmetric solutions
i.e. set $\Lambda(0)=0$ and choose $\lambda=1.5$ and  $\mu~\in~
\{1,0.5,-0.5\}$. The resulting solution curves are shown in 
Fig. \ref{figure1}. In a second run we maintain $\Lambda(0)=0$ but choose
$(\lambda=-0.5,\mu=0.5),~(\lambda=-1,\mu=-0.25)~\textrm{and}~(\lambda=-1,\mu=0.5)$
and the graphs are shown in Fig. \ref{figure2}.

\begin{figure}[ht]
\includegraphics[width=8cm]{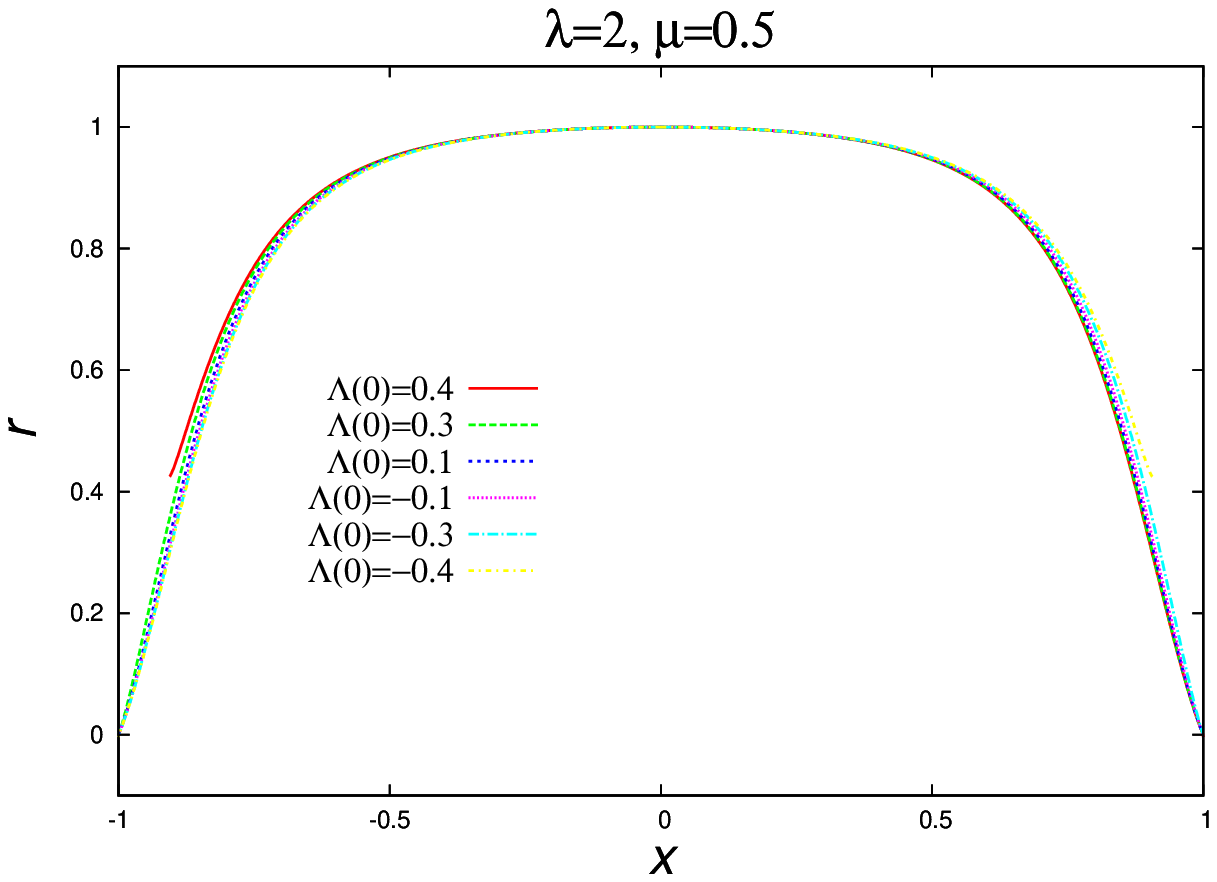}
\includegraphics[width=8cm]{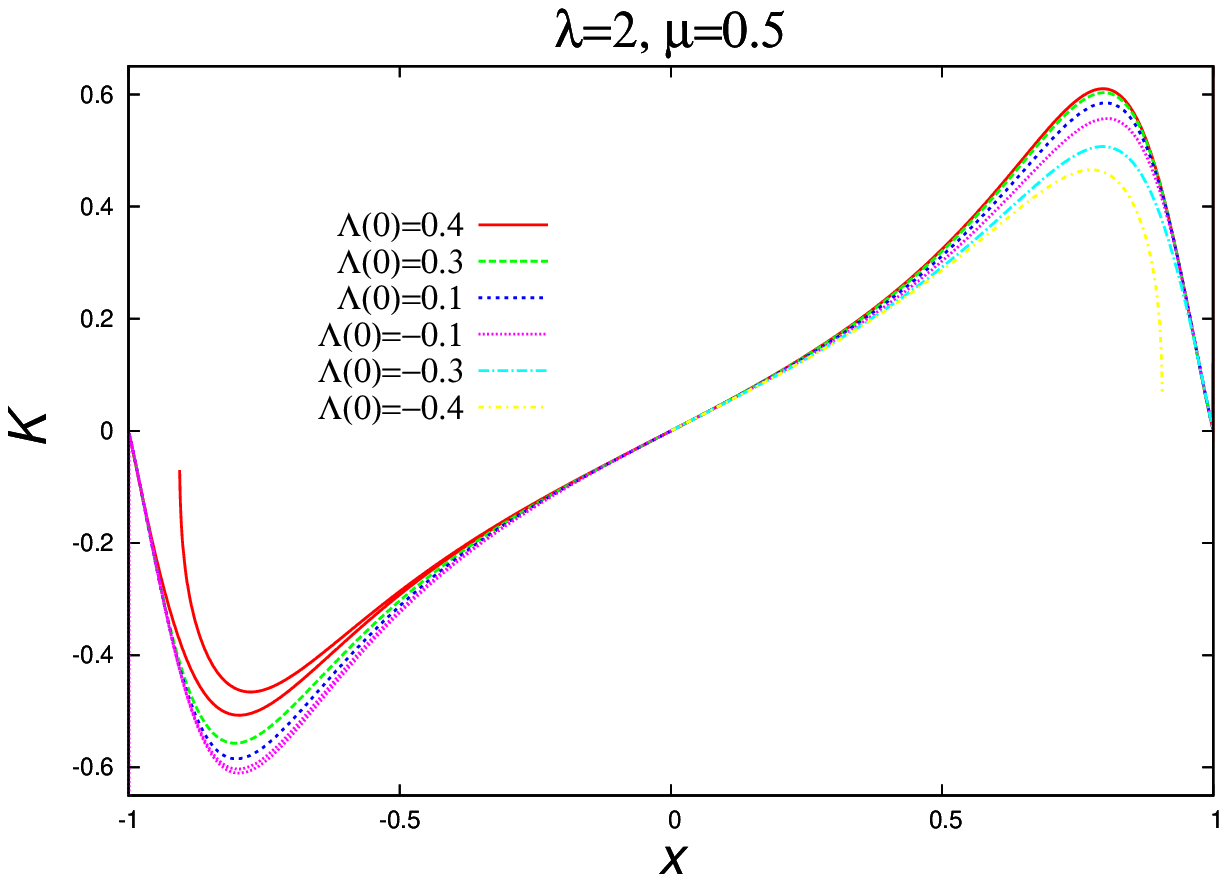}
\includegraphics[width=8cm]{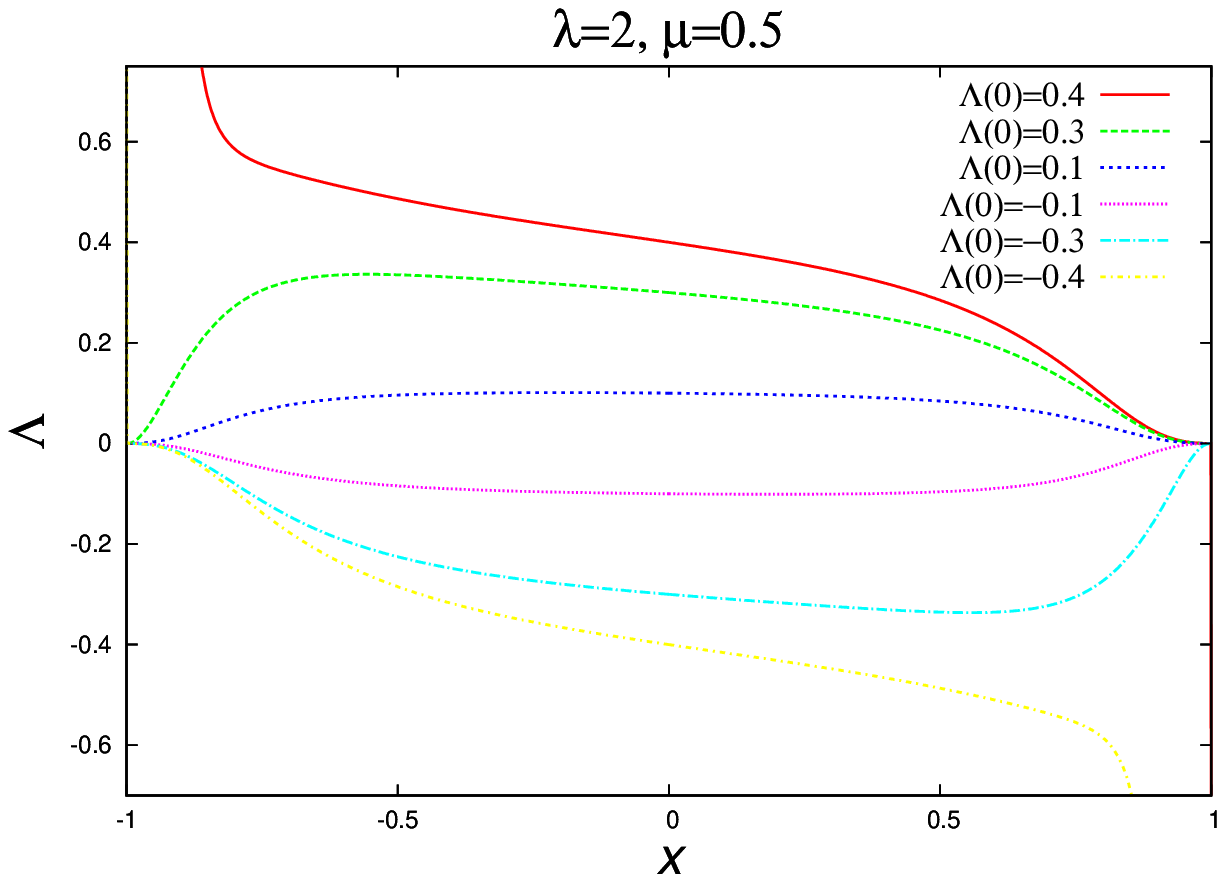}
\includegraphics[width=8cm]{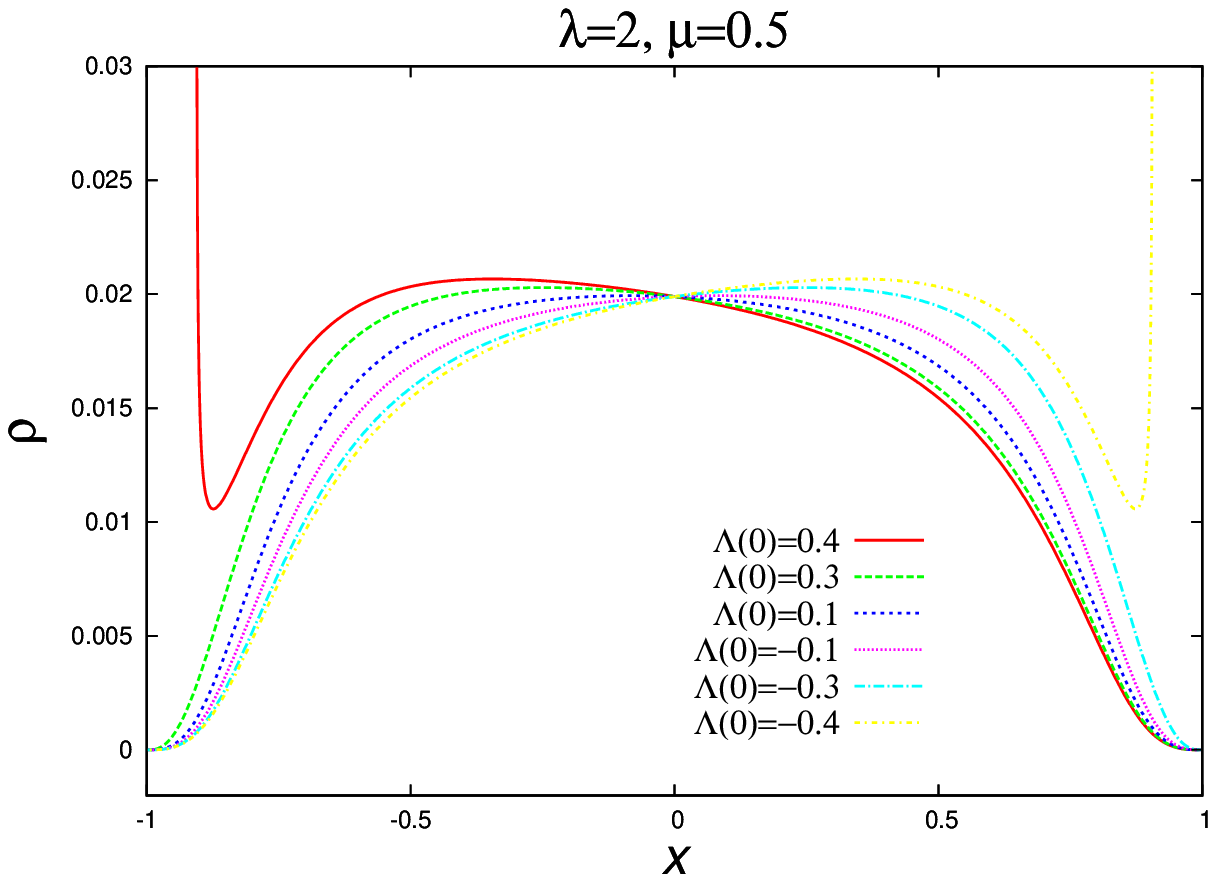}
\caption{In these graphs we have taken $\lambda=2,~\mu=0.5$ so
that $1-\lambda+2\mu=0$, and we have kept them constants throughout.
We gradually change $\Lambda(0)$ starting from
$\Lambda(0)=0$ and increasing (decreasing)  towards positive
(negative) values. Notice that after a characteristic value of
$\Lambda(0)$ the solutions become unbounded at one end.
\label{figure3}}
\end{figure}

\begin{figure}[ht]
\includegraphics[width=8cm]{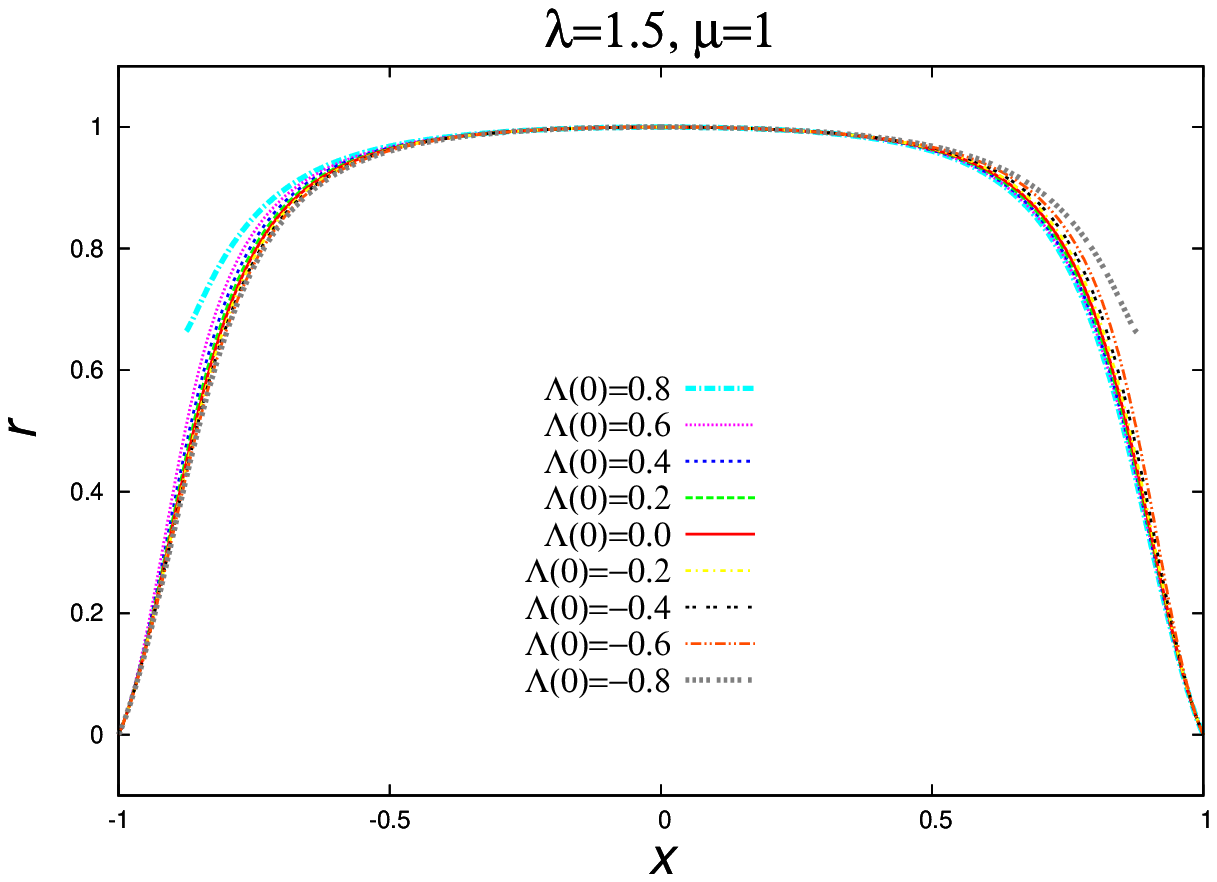}
\includegraphics[width=8cm]{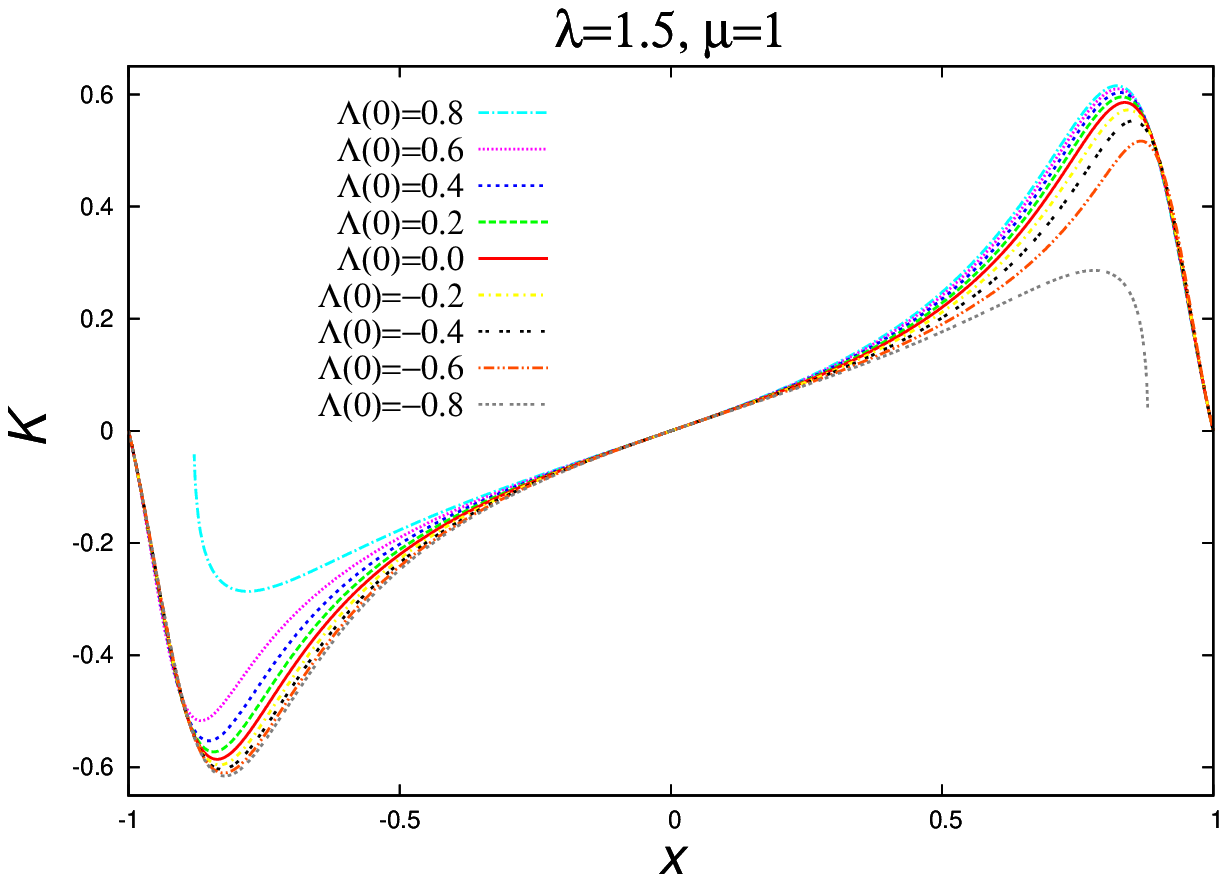}
\includegraphics[width=8cm]{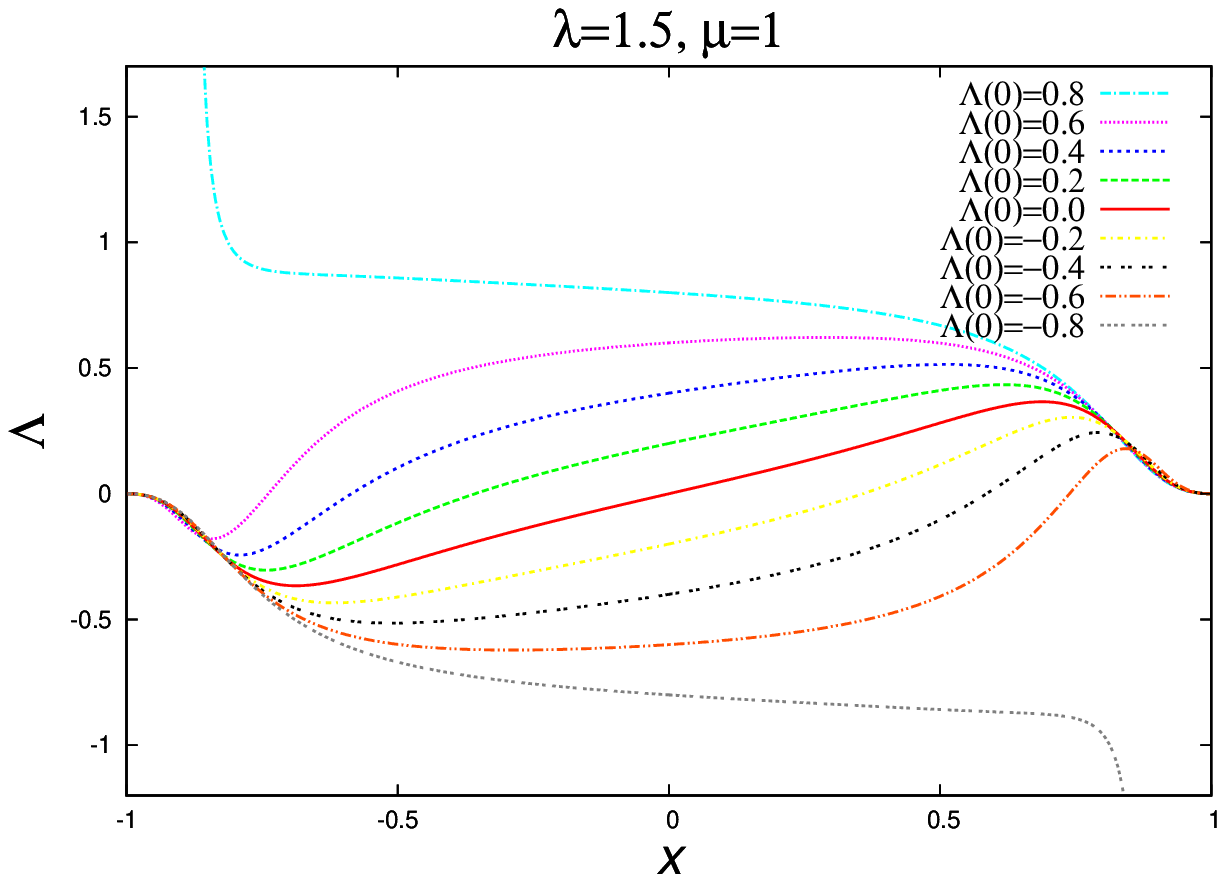}
\includegraphics[width=8cm]{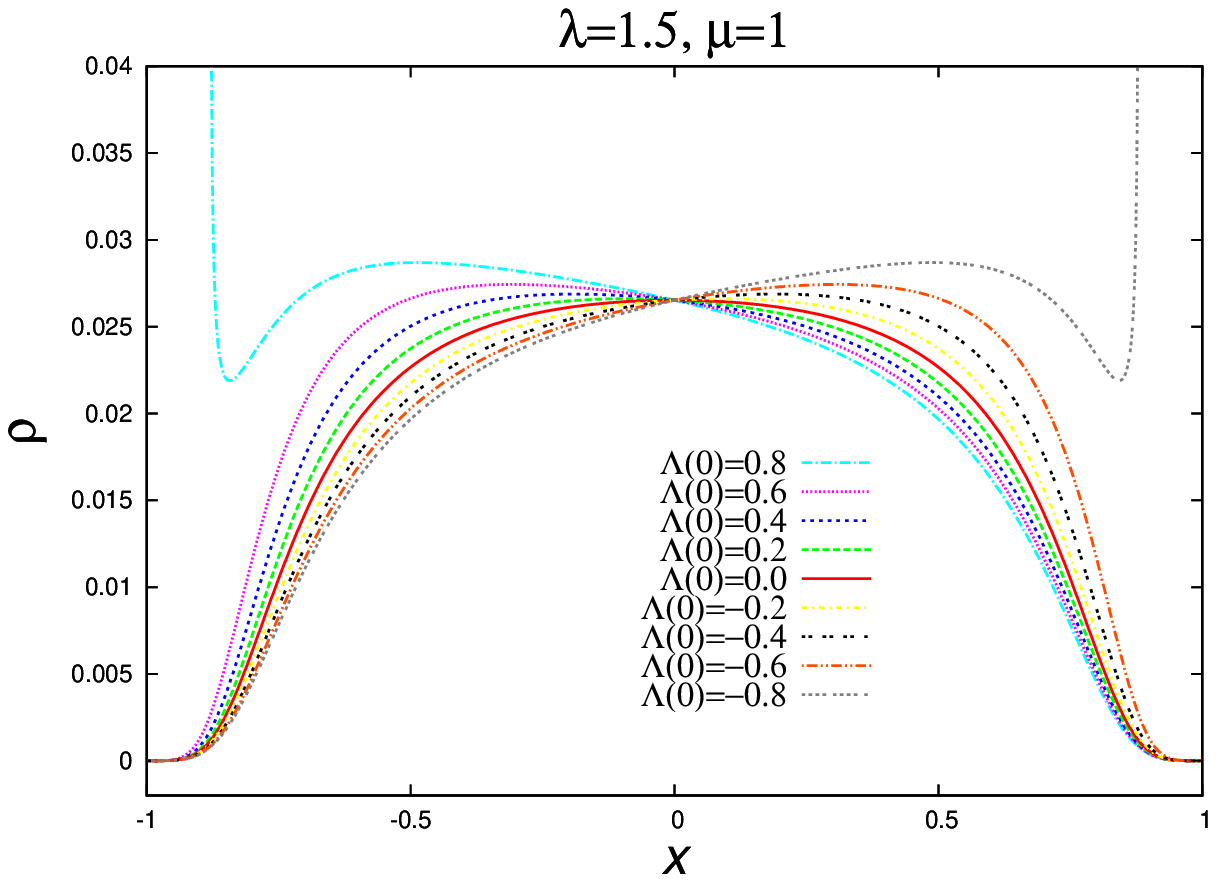}
\caption{In the graphs we have taken $\lambda=1.5,~\mu=1$. 
Like in Fig.(\ref{figure3}), we vary
$\Lambda(0)$ from  $\Lambda(0)=0$ up to $\Lambda(0)=0.8$ and $\Lambda(0)=-0.8$.
Again after some characteristic value of $\Lambda(0)$
the solutions become diverging at one end.
\label{figure4}}
\end{figure}

\begin{figure}[ht]
\includegraphics[width=8cm]{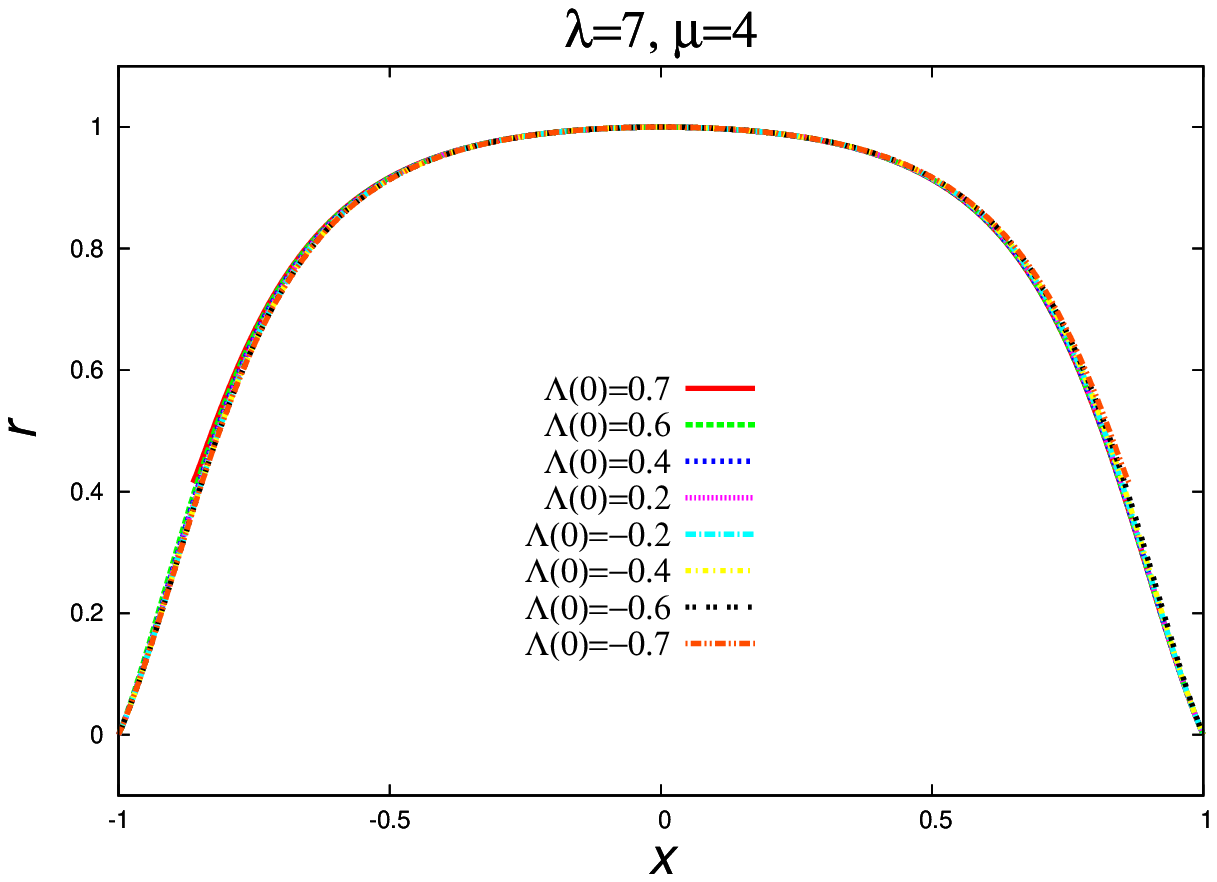}
\includegraphics[width=8cm]{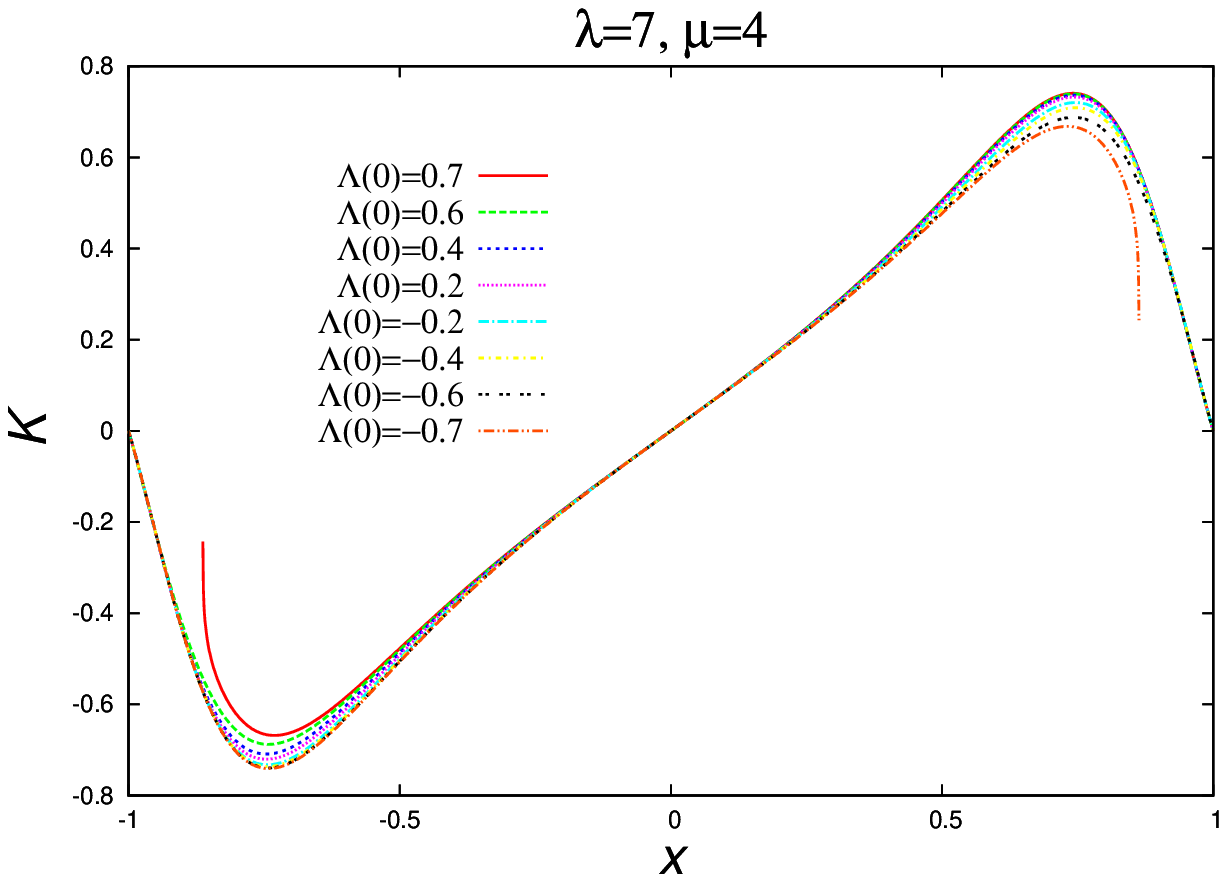}
\includegraphics[width=8cm]{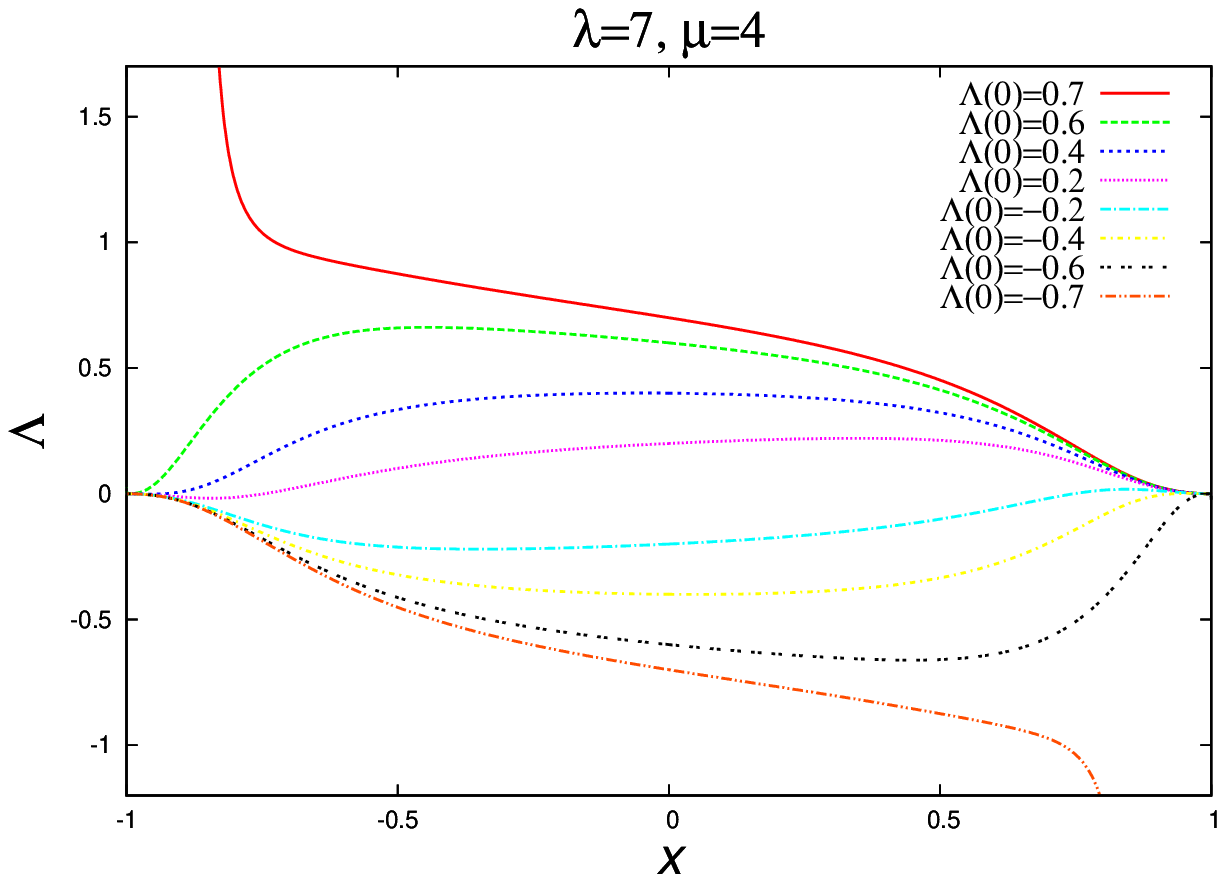}
\includegraphics[width=8cm]{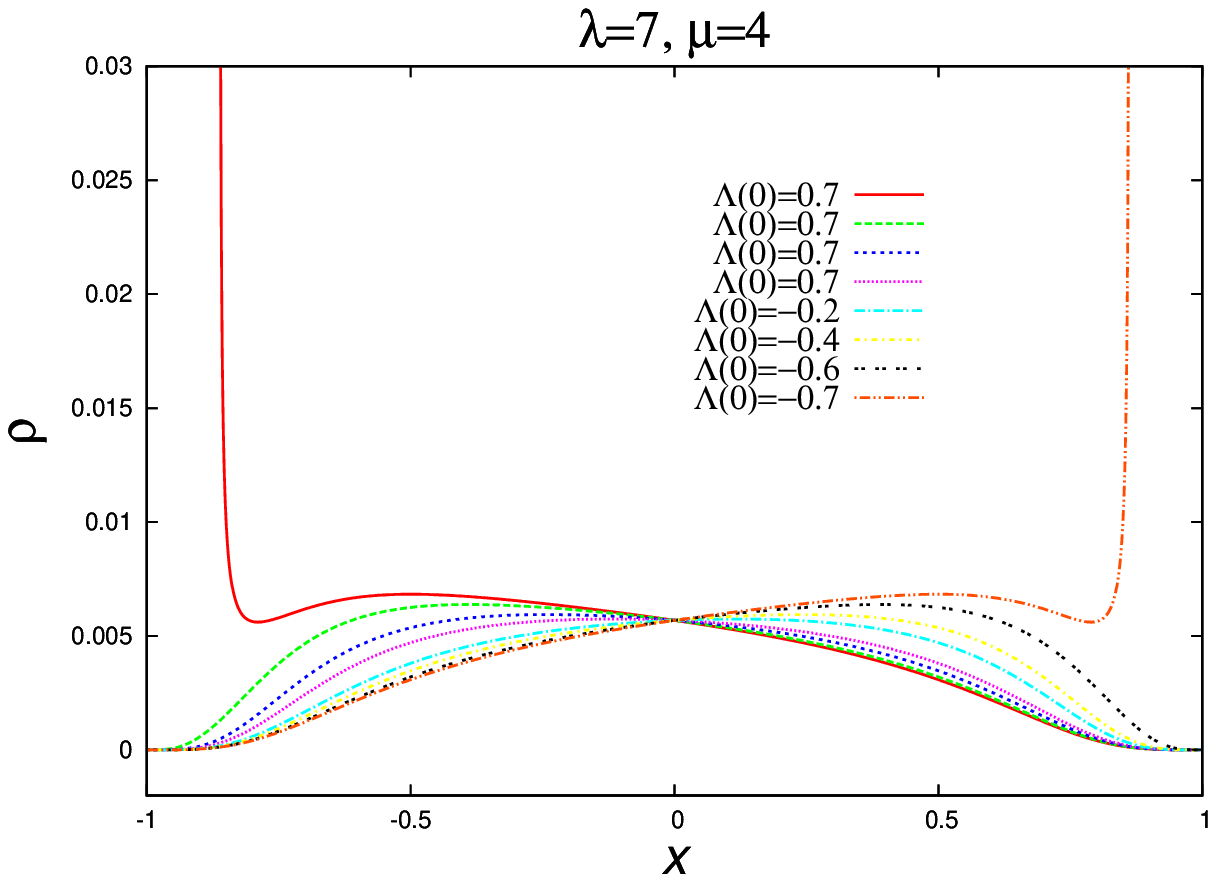}
\caption{We repeat the same analysis as in Figs. (\ref{figure3}, \ref{figure4}) except that we have chosen
$\lambda=7,~\mu=4$. Again, the solutions diverge after some
characteristic value of $\Lambda(0)$. \label{figure5}}
\end{figure}

A number of runs aims to get insights in the behavior of reflectionally
symmetric solutions upon changing the value of $\Lambda(0)$. For
these runs we consider a reflectionally symmetric solution for
fixed $(\lambda, \mu)$,
and vary $\Lambda(0)$ from
$\Lambda(0)=0$ towards positive and negative values.
We have used three  different values for $(\lambda, \mu)$ and the results of
these runs are shown in Figs. \ref{figure3},\ref{figure4},\ref{figure5}.

\begin{table}[ht]
\begin{center}\begin{tabular}{|c|c|c|l|}
\hline
Case    & $\lambda$     &  $\mu$        & Behavior\\
\hline
1       & 1.5           & 1             & Asym. Flat  \\
2       & 1.5           & 0.5           & Decaying Non - Asym.Flat\\
3       & 1.5           & -0.5          & Non - Asym.Flat\\
4       & -0.5          & 0.5           & Non - Asym.Flat\\
5       & -1            & -0.25         & Decaying Non - Asym.Flat\\
6       & -1            & -0.5          & Asym.Flat \\
\hline
\end{tabular}\end{center}
\caption{Table showing the values of the parameters $(\lambda, \mu)$ used in 
Figs. \ref{figure1},\ref{figure2}.}
\end{table}

\section{Existence of Asymptotically flat wormholes\label{IV}}

\noindent The numerical outputs displayed in Figs. \ref{figure1}-\ref{figure5}
show that solutions are
divided into two families: the first one contains solutions where all
variables decay to zero as $x\to\pm\ 1$ whereas the second one
contains solutions where the variables become unbounded in one (or
both) ends. This behavior raises a number of questions: Which if any,
of the decaying to zero solutions describe asymptotically flat
wormholes? How the initial conditions and values of $(\lambda,\mu)$
affect the global behavior of the solutions? Bellow, we provide an
explanation of that behavior and draw a few conclusions
based on these graphs.

We begin by first discussing an exact solution of
(\ref{a3}-\ref{a7}). This solutions is generated by choosing
$(\lambda,\mu)$ so that $1-\lambda+2\mu=0$ and taking $\Lambda(l)=0$
as the global solution of eq. (\ref{a5}). The restriction
$1-\lambda+2\mu=0$ implies that the EOS takes the form:

\begin{eqnarray}
 \tau=\lambda\rho c^{2},~P=\frac{\lambda-1}{2}\rho c^{2},~
 \lambda>1~\textrm{or}~\lambda<0.
\end{eqnarray}

\noindent and the resulting solution in
 curvature coordinates is described by (for details see \cite{JGonzalez-FGuzman-TZannias}):

\begin{eqnarray}
 \textbf{g}&=&-dt^{2}
+\frac{dr^{2}}{1-\left(\frac{r_{0}}{r}\right)^{\frac{\lambda-1}{\lambda}}}
+r^{2}d\Omega^{2},
~r~\in~(r_{0},\infty),\label{ddd3}\\
\hat{k}\rho(r)c^{2}&=&
\frac{1}{r^{2}_{0}\lambda}\left(\frac{r_{0}}{r}\right)^{\frac{3\lambda-1}{\lambda}},
~r~\in~(r_{0},\infty).\label{ddd4}
\end{eqnarray}

\noindent For latter use, we notice  that for $\lambda>1$, the density
decays according to
$\hat{k}\rho(r)c^{2}=O(r^{-(2+\epsilon)}),~0<\epsilon<1$, while for
$\lambda<0$, $\hat{k}\rho c^{2}=O(r^{-(3+\epsilon)}),~\epsilon>0$.

\noindent In the remaining part of this section we analyze the
asymptotic behavior of the solutions. We consider a solution that decays to zero as 
$l\to\pm\infty$ (the graphs in Figs. \ref{figure1}-\ref{figure5} show that such solutions exist).
For such solution and away from the throat we introduce
coordinates $(t,r,\theta,\phi)$ so that:

\begin{eqnarray}
 \textbf{g}&=&-e^{2\Phi(r)}dt^{2}+\frac{dr^{2}}{1-\frac{2m(r)}{r}}+r^{2}d\Omega^{2},~~r >r_{0}=r(0)>0,\\
m(r)&=&\frac{r_{0}}{2}+\frac{\hat{k}c^{2}}{2}\int_{r_{0}}^{r}r^{'~2}\rho(r')dr'.
\end{eqnarray}

Either from (\ref{a3}-\ref{a6}) or directly from
$G_{\alpha\beta}=\hat{k}T_{\alpha\beta}$ we find:

\begin{eqnarray}
&&\frac{dm(r)}{dr}=\frac{\hat{k}c^{2}}{2}r^{2}\hat{\rho}(r),\,
\,\,\,r\,\,\,\in\,\,\,(r_{0},\infty),\label{e}\\
&&\frac{d\tau(r)}{dr}=\left[\rho(r)
c^2-\tau(r)\right]\frac{d\Phi(r)}{dr}-\frac{2[P(r)+\tau(r)]}{r},~~r~\in~(r_{0},\infty),
\label{e1}\\
&&\frac{d\Phi(r)}{dr}=\frac{-\hat{k}\tau(r)r^{3}
+2m(r)}{2r(r-2m(r))},\,\,\, r\,\,\,\in\,\,\,\,(r_{0},
\infty).\label{e2}
\end{eqnarray}

\noindent Taking into account the EOS and introducing the variables:

\begin{eqnarray}
 W(r)=\frac{2m(r)}{r},\,\,\,\ R(r)=\hat{k}\rho(r)c^{2}r^{2},\label{ttzz}
\end{eqnarray}

\noindent (\ref{e}-\ref{e2}) become:

\begin{eqnarray}
r\frac{dW(r)}{dr}&=&-W(r)+R(r),\label{d1}\\
r\frac{dR(r)}{dr}&=&-\frac{2\mu}{\lambda}R(r)+\left(\frac{1}{\lambda}
-1\right)\frac{-\lambda
R(r)+W(r)}{2(1-W(r))}R(r),\label{d2}\\
r\frac{d\Phi(r)}{dr}&=&\frac{-\lambda
R(r)+W(r)}{2(1-W(r))}.\label{d3}
\end{eqnarray}

\noindent This system becomes singular at three places: as $r\to r_0$, on any $r \in
(r_0, \infty)$ so that $W(r)=1$, and as $r\to\infty$.  Since for our
background solution all fields decay to zero as $l\to\pm\infty$, we
assume that (\ref{d1}-\ref{d3}) is defined in a domain
$[R_{0},\infty)$ where $R_{0}$ is sufficiently large so that:

\begin{eqnarray}
 W(r)<1,~R(r)<1~,\forall~~~r~\in~ [R_{0},\infty).\label{B11}
\end{eqnarray}

\noindent and additionally: $\lim_{r\to\infty}
W(r)=0,~\lim_{r\to\infty}R(r)=0$. On $[R_{0},\infty)$ we introduce a
new variable $t$ via:

\begin{eqnarray}
r=r_{0}e^{t}, t~\in~[T,\infty),~~T=\log
\left(\frac{R_{0}}{r_{0}}\right),
\end{eqnarray}

\noindent
so that (\ref{d1}-\ref{d3}) becomes an autonomous system:

\begin{eqnarray}
\frac{dW(t)}{dt}&=&-W(t)+R(t),\label{f8}\\
\frac{dR(t)}{dt}&=&-\frac{2\mu}{\lambda}R(t)+\left(\frac{1}{\lambda}-1\right)\frac{-\lambda
R(t)+W(t)}{2(1-W(t))}R(t),\label{f9}\\
\frac{d\Phi(t)}{dt}&=&\frac{-\lambda
R(t)+W(t)}{2(1-W(t))}.\label{f10}
\end{eqnarray}

\noindent Since solutions of (\ref{f10}) are determined by
$(W(t),R(t))$, we restrict our attention to the first two equations.
They can be written in the form:

\begin{eqnarray}
 \frac{d\textbf{x}(t)}{dt}=\Lambda\textbf{x}(t)+
\textbf{F}(\textbf{x}(t)),~t~\in~[T,\infty),
\end{eqnarray}

\noindent where $\textbf{x}(t)=(W(t),R(t))^{T}$, $\Lambda$ stands for the
matrix:

\begin{eqnarray}
\Lambda=\left(
\begin{array}{cc}
-1 & 1\\
0 & -\frac{2\mu}{\lambda}
\end{array}
\right),
\end{eqnarray}

\noindent while $\textbf{F}$ is defined by:

\begin{eqnarray}
 \textbf{F}:A\subset I\!\!R^{2}\to I\!\!R^{2}:(\hat{W},\hat{R})\to
\textbf{F}(\hat{W},\hat{R})=0, ~~(\frac{1}{\lambda}-1) \frac{\hat{W}
-\lambda\hat{R}}{2(1-\hat{W})}\hat{R}.
\end{eqnarray}

\noindent According to (\ref{B11}), we take:

\begin{eqnarray}
 A=\left\{\textbf{x}=(\hat{R},\hat{W}) ~~|~~~  |\textbf{x}|\leq R<1 \right\},
\end{eqnarray}

\noindent where we employ:
 $|\textbf{x}|=
|x^{1}|+|x^{2}|$ as the norm of $I\!\!R^{2}$,
 \cite{XX}. On this $A$, it is easily seen that $\textbf{F}$
is $C^{\infty}$ actually analytic, and  moreover it satisfies:

\begin{eqnarray}
 |\textbf{F}(\textbf{x})|\leq
\left(M\frac{(1-\lambda)^{2}}{|\lambda|}~|\textbf{x}|\right)|\textbf{x}|\leq
N |\textbf{x}|,\label{a12}
\end{eqnarray}

\noindent where

\begin{displaymath}
 M=\textrm{max}_{~~W~\in~ A}~~|[2(1-\hat{W})]^{-1}|
~~\textrm{and}~~N=\textrm{max}_{~x~\in~A}\left(M\frac{(1-\lambda)^{2}}{|\lambda|}
 |\textbf{x}|\right).
\end{displaymath}

\noindent Let now the IVP:

\begin{eqnarray}
 \frac{d\textbf{x}(t)}{dt}&=&\Lambda\textbf{x}(t)+\textbf{F}(
\textbf{x}(t)),~t~\in~[T,\infty),\label{h1}\\
\textbf{x}(T)&=&\textbf{x}_{0},~~\textbf{x}_{0}~~ \in
~~A.\label{f2}
\end{eqnarray}

\noindent By the Picard-Lindelof's theorem \cite{Picard}, there
exist unique solutions $\textbf{x}(t)$ defined on
$[T,\epsilon),~~\epsilon>0$, and by the variation of constants formula
\cite{YY}  this solution is described by:

\begin{eqnarray}
 \textbf{x}(t)=e^{(t-T)\Lambda}\textbf{x}_{0}
+\int_{T}^{t}e^{(t-s)\Lambda} \textbf{F}(\textbf{x}(s))ds,~~
t~\in~[T,\epsilon).\label{f1}
\end{eqnarray}

\noindent Although this $\textbf{x}(t)$ is defined only on
$[T,\epsilon),~~\epsilon>0,$ as long as $|\textbf{x}_{0}|$ is small
enough and under some conditions upon the eigenvalue
$\lambda_{1}=-1,~ \lambda_{2}=-\frac{2\mu}{\lambda}$ of $\Lambda,$
it can be continued so that remains in $A$ $\forall
~t~\in~[T,\infty)$. For that, let us suppose $(\lambda,\mu)$ are
chosen so that $\lambda\mu>0$. This restriction implies
$\lambda_{2}=-\frac{2\mu}{\lambda}<0$  and thus both eigenvalues of
$\Lambda$ are negative. Standard estimates \cite{Picard,YY} show that there exist $K>0$
and $\rho>\textrm{max}(\lambda_{1},\lambda_{2})$ such that:

\begin{eqnarray}
|e^{(t-T)\Lambda}|\leq Ke^{\rho(t-T)}~~\forall~~t\geq T\label{B44}.
\end{eqnarray}

\noindent As long as  $\textrm{max} (\lambda_{1},\lambda_{2})<0$, we can
always choose a $\rho$ so that: $\textrm{max}
(\lambda_{1},\lambda_{2}) <-|\rho|<0$. For such $\rho$ we obtain from
(\ref{f1}) and (\ref{B44}):

\begin{eqnarray}
 |\textbf{x}(t)|\leq K e^{-|\rho|(t-T)}|\textbf{x}_{0}|
+K\int_{T}^{t}e^{-|\rho|(t-s)}|\textbf{F}
(\hat{\textbf{x}}(s))|ds.
\end{eqnarray}

\noindent Upon multiplying both sides by $e^{|\rho| t}$ we arrive
at:

\begin{eqnarray}
e^{|\rho| t}|\textbf{x}(t)|\leq K
e^{|\rho| T}|\textbf{x}_{0}|+KN\int_{T}^{t}e^{|\rho|s}|\textbf{x}(s)|ds,
\end{eqnarray}

\noindent where we made use  of (\ref{a12}). By appealing to
Gronwall inequality \cite{Gronwall} we conclude that:

\begin{eqnarray}
|\textbf{x}(t)|\leq
K|\textbf{x}_{0}|e^{-(t-T)[|\rho|-KN]},~t>T,\label{a13}
\end{eqnarray}

\noindent and by shrinking the size of $A$
if necessary, we can always make $|\rho|-KN>0$ and thus the solution
$\textbf{x}(t)$ of (\ref{h1},\ref{f2}) remains in $A$,  $\forall~
t>T$.

 This analysis shows that as long as $\lambda\mu>0$, the
solutions to the IVP (\ref{h1},\ref{f2}) are dominated by the linear 
part and since both eigenvalues $(\lambda_{1},\lambda_{2})$ are
negative, they decay exponentially to zero as $t\to\infty$. To get
more insights on their behavior we consider:

\begin{eqnarray}
 W(r)=\frac{c}{r}+\frac{c_{1}}{r^{1+\epsilon}},~~\epsilon>0,
\end{eqnarray}

\noindent and returning to (\ref{ttzz}-\ref{d3}) we find:

\begin{eqnarray}
 W(r)&=&\frac{c}{r}+
\frac{c_{1}}{r^{\frac{2\mu}{\lambda}}},\\
\rho(r)&=&
\frac{\rho_{0}}
{r^{2+
\frac{2\mu}{\lambda}}}
+O\left(\frac{1}{r^{3}}\right),~ \rho_{0}=-(\frac{2\mu}{\lambda}-1)c_{1},\\
\Phi(r)&=&\Phi_{0}-\frac{c}{2r}-\frac{c_{1}\lambda(1+2\mu-\lambda)}{4\mu}\frac{1}{r^{2\frac{\mu}{\lambda}}}
+O(\frac{1}{r^{2}}).
\end{eqnarray}

\noindent  Always under the assumption $\lambda\mu>0$, if
$\lambda_{2}<\lambda_{1}=-1$ then
$\frac{2\mu}{\lambda}=1+\epsilon,~\epsilon>0$ and thus:

\begin{eqnarray}
W(r)&=&\frac{c}{r}+\frac{c_{1}}{r^{1+\epsilon}},\label{T1}\\
\rho(r)&=&-\frac{\left(\frac{2\mu}{\lambda}-1\right)c_{1}}{r^{2+\frac{2\mu}{\lambda}}}=-\frac{\epsilon
c_{1}}{r^{3+\epsilon}},~ \epsilon>0,\label{T9}\\
\Phi(r)&=&\Phi_{0}-\frac{c}{2r}-
\frac{c_{1}\lambda(1+2\mu-\lambda)}{4\mu}\frac{1}{r^{1+\epsilon}},~\epsilon>0.\label{T2}
\end{eqnarray}

\noindent If $\lambda_{1}<\lambda_{2}<0$ then
$\frac{2\mu}{\lambda}=1-\epsilon,~\epsilon>0$ and  the solutions
exhibit slower decay rates:

\begin{eqnarray}
W(r)&=&\frac{c}{r}+\frac{c_{1}}{r^{1-\epsilon}},~ \epsilon>0,\label{T3}\\
\rho(r)&=&\frac{\rho_{0}}{r^{2+\frac{2\mu}{\lambda}}}=\frac{\epsilon c_{1}}{r^{3-\epsilon}},~\epsilon>0,\label{T10}\\
\Phi&=&\Phi_{0}-\frac{c}{2r}-\frac{c_{1}(1+2\mu-\lambda)}{4\mu}\frac{1}{r^{1-\epsilon}},~\epsilon>0.\label{T4}
\end{eqnarray}

\noindent It is interesting to note that for the particular case
where $1-\lambda+2\mu=0$, we obtain
$\lambda_{2}=-\frac{2\mu}{\lambda}=-1+\frac{1}{\lambda}$ and thus if
$\lambda>1$ then $\lambda_{1}<\lambda_{2}<0$, which implies
$\hat{k}\rho c^{2}=O(r^{-(2+\epsilon)})$, while for $\lambda<0$ 
$\lambda_{2}<\lambda_{1}$, and thus $\hat{k}\rho
c^{2}=O(r^{-(3+\epsilon)})$. This behavior is in agreement with the
exact solution described by (\ref{ddd3},\ref{ddd4}).

There is an important difference between the  asymptotic behavior
of solutions described by (\ref{T1}-\ref{T2})
and those described by (\ref{T3}-\ref{T4}). The first family has finite ADM mass $M_{ADM}$ while 
for the second family this mass is actually divergent. In order to see that, we notice that the 
spatial metric 
$^{(3)}\gamma$ can be written in the form:

\begin{eqnarray}
 ^{(3)}\gamma=\frac{dr^{2}}{1-W(r)}+r^{2}d\Omega^{2},~r~\in~(R_{0},\infty),\label{T7}
\end{eqnarray}
              
and by employing conformal coordinates takes the form:
\begin{eqnarray}
^{(3)}\gamma=\hat{\Omega}^{2}(R)[dR^{2}+R^{2}d\Omega^{2}]=\hat{\Omega}^{2}(x)[dx^{2}+dy^{2}+dz^{2}],
\label{T6}
\end{eqnarray}

\noindent where

\begin{eqnarray}
 \hat{\Omega}^{2}=\left(\frac{dr(R)}{dR}\right)^{2}\frac{1}{1-W(r)}.\label{T5}
\end{eqnarray}

\noindent In order to evaluate the $M_{ADM}$,
we employ the representation \cite{Murchadha}:

\begin{eqnarray}
M_{ADM}&=&\frac{1}{16\pi}\int_{\Sigma}[^{3}R(\gamma)+
\frac{1}{4}\gamma^{mn}\gamma^{ab}\gamma^{cd}(2\gamma_{mn,d}
\gamma_{ac,b} -2\gamma_{ma,d}\gamma_{nc,b}\label{m8}
\\
\nonumber
 &&+\gamma_{dm,b}\gamma_{nc,a}
-\gamma_{cd,a}\gamma_{mn,b})]\sqrt{\gamma}d^{3}x,
\end{eqnarray}

where $\Sigma$ stands for  any asymptotic end defined by restricting
$r$ so that:  $r>R_{0}$.

\begin{figure}[ht]
\includegraphics[width=8cm]{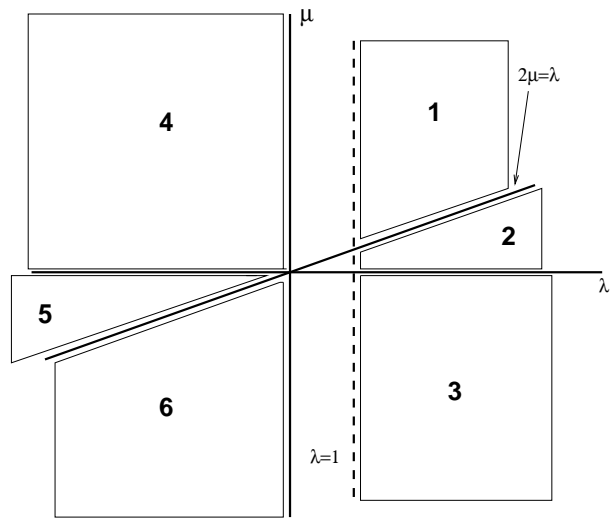}
\caption{Sketch of the parameter space $( \lambda,\mu)$. The regions
labeled by $(3,4)$ do not generate solutions that decay to zero
asymptotically. Regions labeled $(1,6)$ generate solutions decaying
to zero at infinity but the decay is slow. Finally regions $(2,5)$
generate solutions that decay sufficiently fast so that
they have finite ADM mass. \label{figure7}}
\end{figure}

\noindent By appealing to the Hamiltonian constraint $^{(3)}R(\gamma)=2\hat{k}\rho c^{2}$ 
and (\ref{T6}), the right hand side of (\ref{m8}) yields:

\begin{eqnarray}
M_{ADM}=\int_{\Sigma}\left[2\hat{k}\rho c^{2}
-\frac{1}{2\hat{\Omega}^{6}(R)}\left(\frac{d\hat{\Omega}^{2}}{dR}\right)^{2}
\right]\sqrt{\gamma}d^{3}x.
\end{eqnarray}

\noindent However it can be easily seen that this integral converges for the solutions
described by (\ref{T1},\ref{T9}) and diverges for those described by (\ref{T3},\ref{T10}) (for more details see also
\cite{MZ1}).


\section{Discussion \label{V}}

\noindent The results of the present paper
can be succinctly summarized by partitioning the $( \lambda,\mu)$ plane according to whether
the eigenvalues $(\lambda_{1},\lambda_{2})$ of the matrix $\Lambda$ satisfy:
$\lambda_{2}<\lambda_{1}$ or $\lambda_{1}<\lambda_{2}<0$ (see Fig. \ref{figure7}). The
values of $(\lambda,\mu)$ required to generate
asymptotically flat solutions in the sense that
the $M_{ADM}$ is finite is indicated.
Our analysis
shows that solutions of (\ref{a}-\ref{eq:final}) subject to
(\ref{C1},\ref{C2}) if they are decaying as $r\to\infty$, then $(\lambda,\mu)$ should lie in the regions $(1,2)$ and
$(5,6)$ of Fig. \ref{figure7} and the decay rates are those described in (\ref{T1}-\ref{T2})
and (\ref{T3}-\ref{T4}).
It is important however to stress a point. Although
the numerical outputs show that there exist initial conditions so that
the solutions
are decaying as $r\to\infty$ and moreover are asymptotically flat whenever $(\lambda,\mu)$
are taken in regions $(2,5)$, we have not shown
that for any $(\lambda,\mu)$, lying
on regions $(2)$ or $(5)$ of Fig. \ref{figure7}, there exist
initial conditions so that the solutions
of (\ref{a}-\ref{eq:final}) are 
reaching the asymptotic region. This for the moment is an open issue.

Our results establish the existence of 
reflectionally symmetric asymptotically flat
wormholes having throat of arbitrary area.
This
conclusion is a consequence of the rescaling property of the solutions
under rescaling of the throat area. As far as non reflectional
wormholes are concerned, the graphs in Figs. \ref{figure4}-\ref{figure5} show that
they do exist (notice for these figures $\lambda_{2}<-1$). However
they do not exist for all values of
$\Lambda(0)$. The numerical outputs show that after some critical values of $\Lambda(0)$ the
solutions become unbounded. This suggests that
$A(0)$ and $\Lambda(0)$ may not be taken independently although
the nature of any constraint of this sort is for the moment unknown
\cite{Note2}.


\acknowledgments

\noindent It  is our pleasure to acknowledge many stimulating
discussion with the members of the relativity group at IFM-UMSNH, 
in particular J. Estevez-Delgado and Olivier Sarbach. This work was
supported in part by grants: CIC-UMSNH 4.7, 4.9 and 4.23, SEP-PROMEP
UMICH-CA-22, UMICH-PTC-210, COECyT Michoac\'an S08-02-28 and E9507, 
CONACyT 79601 and 79995.



\begin{thebibliography}{99}
\bibitem{N2} Although we are ignorant of the microphysics describing
 exotic matter and thus we cannot justify
the choice for the EOS employed in the present work, our aim is to
address the global existence of spherical wormholes away from the
usual approach where one first specifies the wormhole metric and
subsequently tailor the structure of exotic matter required to
support it.
\bibitem{note1} Ought to be stressed that by utilizing
eq. (\ref{a7}) one of the four variables can be expressed in terms 
of the other three. For instance: 
$\lambda\hat{k}\rho(l) c^{2}=\hat{r}^{2}(l)-
\frac{K^{2}(l)}{4}-\Lambda(l)K(l)$ and upon subsituting this expression into eq.
(\ref{a6}) we find that it is identically satisfied by virtue of (\ref{a3},\ref{a4})
and (\ref{a5}). Thus the system (\ref{a3}-\ref{a7}) 
is a system of three eqs for three unknowns.
As such the initial
condition for $\rho(0)$ cannot be chosen
independently, a point that is clearly reflected in the initial conditions described in eqs.
 (\ref{X},\ref{C1}) and 
(\ref{C2}). Since we resort into grafical representations
of the solutions, we shall consider (\ref{a3}-\ref{a7}) as a system of four eqs. 
for four variables.
In this way the numerical plots exhibit the behavior of the
geometrical as well as the matter variables.
\bibitem{N3} In a cosmological setting, 
a phantom configurations is restricted by the requirement
that observeres comoving
with the cosmological expansion measure $\rho c^{2}>0$. The choice
$(\rho(0)<0,\lambda<0)$ is consistent with the wormhole initial
conditions and thus we shall analyze this case as well. It may be
worth recalling that massless $K$-essence belongs to this latter
family of exotic matter.
\bibitem{XXX} This property of solutions is a generic property of
the system (see for instance discussion in \cite{MZ2}).
\bibitem{Gronwall} We recall that this inequality states that if  $K\geq 0$ and $(f,g)$
are non negative functions in $C[a,b]$ satisfying:
         $f(t)\leq K + \int_{\alpha}^{t}f(s)g(s)ds$, $t\in[a,b]$  then
         $f(t)\leq Ke^{\int_{\alpha}^{t}g(s)ds}$ (see for instance \cite{Picard,YY}).
\end{thebibliography}
\end{document}